\documentclass[11pt,twoside,a4paper]{cernyrep}

\usepackage{lineno}
\usepackage{graphicx}%
\usepackage{multirow}%
\usepackage{amsmath,amssymb,amsfonts}%
\usepackage{amsthm}%
\usepackage{mathrsfs}%
\usepackage[title]{appendix}%
\usepackage[table,xcdraw]{xcolor}
\usepackage{textcomp}%
\usepackage{manyfoot}%
\usepackage{booktabs}%
\usepackage{algorithm}%
\usepackage{algorithmicx}%
\usepackage{algpseudocode}%
\usepackage{listings}%
\usepackage{placeins}
\usepackage{subfig}
\usepackage{subfiles}%
\usepackage{textgreek}
\usepackage{comment}
\usepackage{siunitx}
\usepackage{tikz} 
\usepackage{tabularx}       
\usepackage{makecell}       
\usepackage{pdflscape} 
\usepackage{booktabs} 
\usepackage{amsmath} 
\usepackage{makecell}
\usepackage{multirow} 
\usepackage{tcolorbox} 
\usepackage[normalem]{ulem}
\usepackage{cancel}
\usepackage{wrapfig}
\usepackage{rotating}
\usepackage{wasysym}   
\usepackage{hyperref}
\hypersetup{%
        colorlinks,
        breaklinks=true,
        plainpages=false,%
        citecolor=blue,
        linkcolor=blue,
        urlcolor=blue,
        bookmarksopen=true,%
        bookmarksnumbered=false,%
        bookmarksdepth=5%
}

\newcommand{\micron}{\ensuremath{\upmu \mathrm{m}}}

\newcommand{\keV}{\si{\kilo\electronvolt}}
\newcommand{\MeV}{\si{\mega\electronvolt}}
\newcommand{\GeV}{\si{\giga\electronvolt}}

\newcommand{\mlab}[1]%
    {\mbox{}\marginpar{\raggedright\hspace{0pt}\tiny {\color{red}{#1}}}}
\newcommand{\red}[1]{\emph{\color{red} #1}}

\newcommand{\as}{\alpha_{\rm S}}

\newcommand{\invab}{{\rm ab}^{-1}}

\usepackage{array}
\usepackage{soul}
\usepackage{enumerate}
\usepackage[referable]{threeparttablex}
\usepackage{footnote}
\usepackage{comment}
\usepackage[acronym,toc]{glossaries}

\begin{document}

\pagestyle{empty}

\vspace{-1 cm}

\begin{center}

\textbf{\LARGE FCC-ee Lessons from SuperKEKB}

\begin{center}
{\Large \it Complementary Contribution to the \\ European Strategy for Particle Physics Update 2025 }
\end{center}

 F.~Zimmermann\footnote{frank.zimmermann@cern.ch}, CERN 

 \end{center}

\begin{center}
{Updated on 17 October 2025}
\end{center}


\begin{center}
{\bf Abstract}
\end{center}

\noindent
SuperKEKB has achieved significantly higher specific luminosity than its predecessor KEKB, and it has proven a much more sustainable machine. It has successfully demonstrated several key design 
elements of FCC-ee. The design luminosity has not yet been reached, however. 
This observation is often (mistakenly) used to put into question the reliability of the FCC-ee design luminosity. 
In this note we review the  accomplishments, challenges and obstacles
of SuperKEKB, and compare these with the FCC-ee design.

\noindent




\pagenumbering{arabic}
\setcounter{page}{1}


\section{Introduction}
In 2021, the KEK management invited an 
International Task Force to contribute to the SuperKEKB beam 
commissioning and operation.  
FCC experts from CERN, CEA, CNRS, DESY, INFN, and other European laboratories 
contributed to this effort.  
European participation in SuperKEKB was, and is, partly 
supported by the European Union's EJADE and EAJADE projects.  
Numerous public presentations on the SuperKEKB performance 
were given at the eeFACT series of ICFA Advanced Beam Dynamics workshops, most recently at eeFACT25 in Tsukuba \cite{eefact25}, and also at the annual IPAC accelerator conferences, e.g.~\cite{ipac24,ipac25}.
In spring 2025, a dedicated mini-workshop on SuperKEKB and 
possible implications for FCC-ee was organised 
in hybrid format at CERN and KEK \cite{skekbfcc}.
Most of the material synthesised in the following was 
presented at the CERN-KEK mini-workshop, at the eeFACT workshop series, or at one of the IPAC conferences. 
A few pictures were generated specifically for this report.

\section{Achievements and Goals}
\label{sec:ag}  
\begin{figure}[!ht]%
\centering
\includegraphics[width=0.9\textwidth]{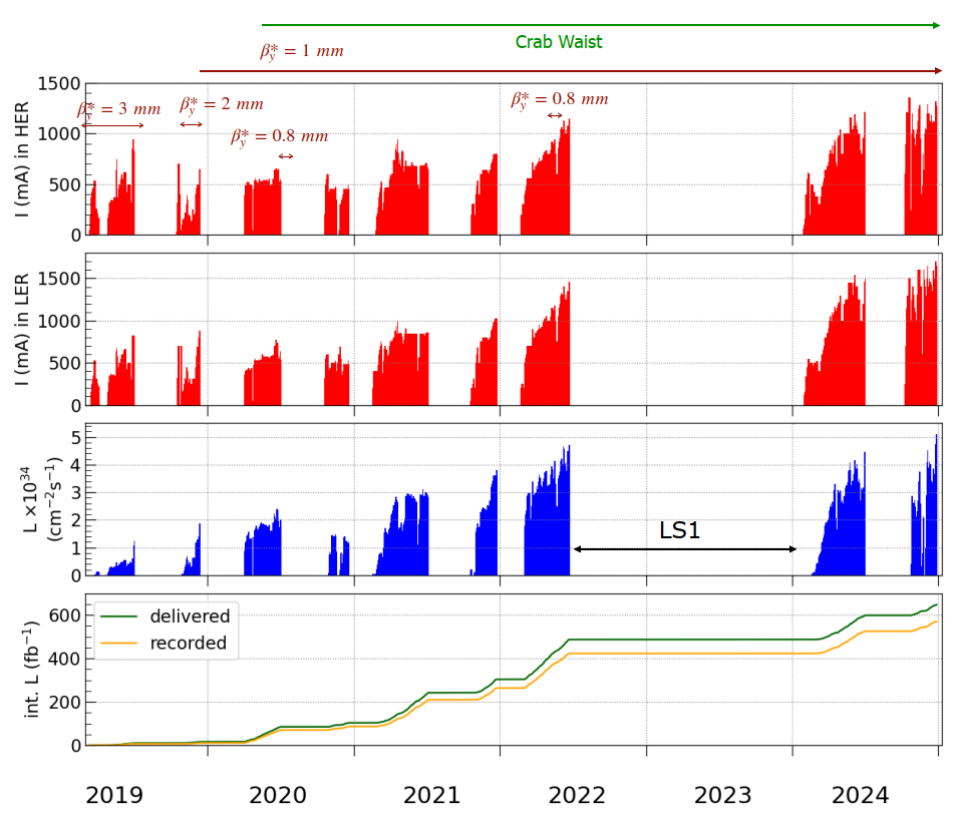}
\caption{Run history of SuperKEKB: beam currents in HER (e$^-$) and LER (e$^+$), 
along with instantaneous and integrated luminosity 
as a function of time from 2019 through 2024, as reported by Y.~Ohnishi \cite{ohnishi25}. }\label{fig:skekbhist}
\end{figure}

The run history of SuperKEKB is illustrated in Fig~\ref{fig:skekbhist}.
We can observe eight rather short runs of typically 3 months duration 
separated by long or very long shutdown
periods. Since it takes 1--2 months to start up the collider, 
this might not be the most efficient mode of operation. 
Regardless, in six of the eight runs, including the 
last one, new luminosity records, which often 
also represented world records,  were established.

Since end of 2019 SuperKEKB has routinely been 
operating with a vertical beta function at the collision point, $\beta_{y}^{\ast}$, 
of  1 mm, which is the design value for the FCC-ee Higgs factory. Occasionally, during short periods the
machine was operated for luminosity production 
with an even smaller  $\beta_{y}^{\ast} =  0.8~$~mm, 
nearly equal the smallest value ever to be 
considered for FCC-ee  (namely,
0.7 mm at the Z pole).

The ``virtual'' crab waist scheme, first developed
for the FCC-ee collider optics in the year 2015 \cite{PhysRevAccelBeams.19.111005}, 
was successfully implemented at SuperKEKB in mid 2020,
and has been used ever since.
In 2024, both SuperKEKB 
rings achieved and operated with beam currents well above the $\sim$1.3 A maximum beam current foreseen for the 
FCC-ee (and needed only at the Z pole).

\renewcommand{\thefootnote}{\roman{footnote}}

In short, SuperKEKB has successfully demonstrated three essential FCC-ee design features:
\begin{itemize}
  \item an ultralow $\beta_y^{\ast}$ of 0.8--1.0 mm \footnote[1]{with about 100 (10,000) times larger vertical geometric emittance for the stored (injected) beam than FCC-ee, 
  and about half the effective focal length},  
  \item the virtual crab waist collision scheme, and
  \item electron and positron beam currents exceeding 1.3 A. 
\end{itemize}

Table \ref{tab:par} compares the  
SuperKEKB beam parameters achieved in 2022 and 2024
with their design values and with   
numbers from the former KEKB.

\begin{table}[htbp]
\caption{ Design values and achieved SuperKEKB (SKB) parameters compared with the luminosity-record parameters of the former KEKB. Shown are the beam energies $E_b$, the horizontal and vertical beta functions at the 
collision point $\beta_{x,y}^{\ast}$, the geometric transverse rms emittances $\varepsilon_{x,y}$, 
the total beam currents $I$, the number of bunches per beam $n_b$, the bunch currents $I_b$, the vertical tune shift parameter estimated simple-mindedly from the luminosity $\xi_y$,
the specific luminosity $L_{\rm sp}$, and the total instantaneous
luminosity $L$.
}
\label{tab:par}
\begin{center}
\begin{tabular}{l|cc|cc|cc|cc} 
 \hline\hline
parameter & 	\multicolumn{2}{|c}{KEKB w Belle} & 
\multicolumn{2}{|c}{SKB 2022b}
 &
 \multicolumn{2}{|c}{SKB 27 Dec.~2024} &
 \multicolumn{2}{|c}{SKB design} \\
 & 	LER	& HER	& LER	& HER& LER & HER &  LER	& HER \\
\hline
$E$ [GeV]& 3.5	& 8	& 4	& 7& 	4 & 	7& 	4& 	7\\
$\beta_x^{\ast}$ [mm] & 1200& 	1200	& 80 & 	80	& 60 & 	60	& 32& 	25\\
$\beta_y^{\ast}$ [mm] & 5.9	& 5.9 & 1.0 &1.0 & 1.0 & 1.0 & 	0.27 & 0.30\\
$\varepsilon_x^{\ast}$ [nm] & 18 & 24& 4.0	& 4.6 & 4.0 & 4.6 & 3.2	& 4.6\\
$\varepsilon_y^{\ast}$ [pm] &  150 &	150 & 	$\sim$50& 
$\sim$50 & 	$\sim$70 	& $\sim$70	& 8.6	& 12.9 \\
$I$ [mA]	& 1640	& 1190 & 	1321& 	1099 & 	1632& 1259 & 	3600	& 2600 \\
$n_b$& \multicolumn{2}{|c}{1584}	& 
\multicolumn{2}{|c}{2249} & 	\multicolumn{2}{|c}{2346} & 
\multicolumn{2}{|c}{2500} \\
$I_b$ [mA] & 	1.04 & 	0.75	& 0.587 & 	0.489 & 	0.696 & 	0.537	& 1.44& 	1.04 \\
$\xi_y$ $^{\ddagger}$ & 	0.098 & 0.059 & 	0.0407 &	0.0279	& 0.036 & 	0.027	& 0.069	& 0.060 \\
$L_{\rm sp}$ [$\mu$b$^{-1}$s$^{-1}$mA$^{-2}$]	& 
\multicolumn{2}{|c}{17.1} & 	
\multicolumn{2}{|c}{71.2}	& 
\multicolumn{2}{|c}{58}	& 
\multicolumn{2}{|c}{214} \\ 
$L$ [$10^{34}$~cm$^{-2}$s$^{-1}$]	& 
\multicolumn{2}{|c}{2.11}	& 
\multicolumn{2}{|c}{4.65} & 	
\multicolumn{2}{|c}{5.1} & 
\multicolumn{2}{|c}{80} \\
\hline\hline
\end{tabular}
\end{center} 
$^\ddagger$The beam-beam parameter is computed without accounting for hourglass effect or geometric factor.

\end{table}

After this brief review of SuperKEKB run history, achievements, and goals,  
the next Section \ref{sec:sbl} discusses
a peculiar phenomenon of 
sudden beam loss effects, 
which have interrupted 
collider operation by  
magnet quenches, damaged collimator jaws and even 
some physics detector components, 
and hindered swift beam-current increases.
The larger-than-expected emittances of both stored and injected beams are examined in Sections \ref{sec:sbe} and \ref{sec:ibe}, respectively.
Section \ref{sec:ie} surveys limitations on 
the injection efficiency, 
Section \ref{sec:IR} major challenges 
in the SuperKEKB interaction region.
Obstacles to further squeezing the vertical beta function at the SuperKEKB 
collision point, $\beta_y^{\ast}$, are illuminated in Section \ref{sec:vb}.
Section \ref{sec:specl} shows that several different effects may conspire  
to greatly reduce the SuperKEKB 
specific luminosity at higher bunch currents. 
Our discussion is summarized and some conclusions 
are drawn in Section \ref{sec:sc}.

The appendices provide further details and present additional issues. 
Appendix \ref{sec:diag} calls for better beam diagnostics, in order to
more adequately control residual optical aberration at the collision point and
the vertical emittance of the stored beams.
This vertical emittance might be limited by a significant and  steadily worsening tunnel deformation, discussed in Appendix \ref{sec:tunnelfloor}.
Physical-aperture limitations due to
the interaction region are scrutinized in Appendix \ref{sec:irap}. Appendix \ref{sec:bd} inspects 
additional beam dynamics known to degrade the specific luminosity of SuperKEKB. 

\section{Sudden Beam Loss}
\label{sec:sbl}
\begin{figure}[!ht]%
\centering
\includegraphics[width=0.85\textwidth]{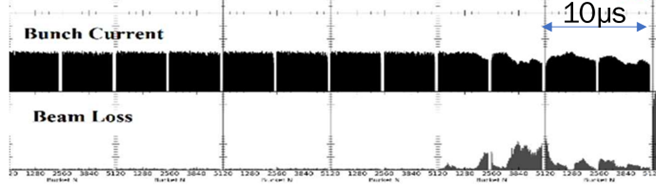}
\caption{An example sudden beam loss event, presented by 
H.~Ikeda \cite{ikeda25}. The revolution period is 10 $\mu$s.
The intensity of two long bunch trains can be seen, turn by turn,  
till the moment of beam abort.
}\label{fig:sblex}
\end{figure}

\begin{figure}[!ht]%
\centering
\includegraphics[width=0.45\textwidth]{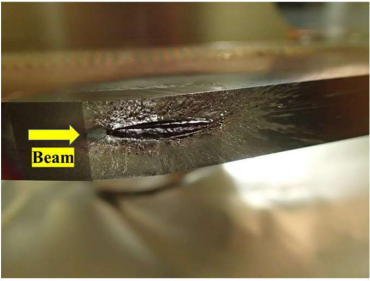}
\caption{A SuperKEKB collimator jaw 
damaged by an SBL event, from 
H.~Ikeda \cite{ikeda25}. 
}\label{fig:sblcoll}
\end{figure}

At SuperKEKB, large sudden beam loss (SBL) events can  occur 
within 1 turn (10 $\mu$s).
An example is shown in Fig.~\ref{fig:sblex}. 
A significant portion of the beam is 
lost in such events before the abort trigger is applied.
These SBLs often lead to damage of collimators (Fig.~\ref{fig:sblcoll}) 
and other  accelerator components, occasionally also 
to quenches of the final focusing 
superconducting magnets (QCS), and to large detector 
backgrounds plus potential damage 
inside the Belle 2 experiment.
SBLs occur for both beams,  in collision and without collision,
but more frequently 
and more catastrophically in the Low Energy positron Ring (LER).
In the LER,  about 250 such events were observed in 2024. 
The rate of SBL events is higher the higher the beam current.
Several reports on SBLs 
were given by H.~Ikeda, e.g.~\cite{ikeda25}.  

\begin{figure}[!ht]%
\centering
\includegraphics[width=0.46\textwidth]{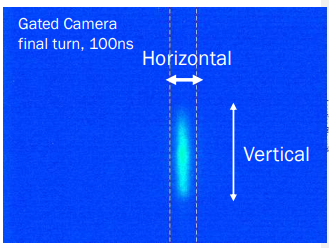}
\includegraphics[width=0.44\textwidth]{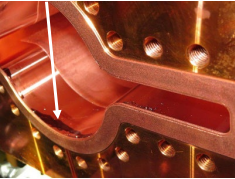}
\caption{Vertical blow during a sudden beam loss event (left), from 
H.~Ikeda \cite{ikeda25}, and black stain next to an ``MO''-type flange (right), shown by K.~Shibata \cite{shibata25}.}\label{fig:sbl}
\end{figure}

The SBLs are often accompanied by local vacuum pressure bursts.
It was possible to induce SBLs by knocking on the vacuum chambers. 
Importantly, there are a few
early warning signs, that could, in principle, be used to trigger an early beam abort. 
Namely, several turns before the sudden beam loss happens, the vertical beam size often increases significantly (Fig.~\ref{fig:sbl}, left picture). 
Also, for certain SBLs, a growing discharge signal from some of the 
clearing electrodes was detected starting 
about ten turns before the sudden beam loss.

At SuperKEKB, a 
vacuum sealant called ``VacSeal'' had been applied to stop vacuum leaks at Matsumoto-Ohtsuka 
(``MO'') type flanges. The MO flanges were specifically developed for SuperKEKB, to mimimise impedance and avoid higher-order modes. 
Many SBLs occurred, along with local 
pressure bursts, near exactly these flanges.
Large black flaky stain, or debris, presumably VacSeal deformed under high heat and 
exposure to synchrotron radiation, is seen inside the vacuum chamber at those MO flanges where VacSeal had been applied (Fig.~\ref{fig:sbl}, right picture). 
Flakes emanating 
from this black stain are thought to be at the origin of the SBLs. 
Indeed, in a few regions were the black stain
was removed by an intervention during the last run, afterwards no further SBLs were observed.
Neither black stain nor SBLs were seen at the locations of standard ``Helicoflex'' gasket flanges, 
even if VacSeal had been used there.
A cleaning campaign for removing all black stain  from inside the vacuum chamber is underway.  

A possible additional source of SBL events 
in SuperKEKB is electric discharges occurring at 
clearing electrodes  \cite{Ikeda25b,sterui} and at ceramic kicker
locations \cite{sterui}.

A model describing the interaction of VacSeal (or other) flakes 
with the circulating beam and explaining the vertical blow up plus beam loss 
has been developed by K.~Ohmi and colleagues \cite{ohmi25b,ohmi2025beamdustinteractionseecollider},
providing a satisfactory explanation for the events observed.

No other accelerator has used MO flanges, and  
no other collider, including PEP-II, KEKB and DA$\Phi$NE, nor any low-emittance light sources, such as ESRF-EBS, APS-U, etc., 
ever reported catastrophic SBL events\footnote{For example, at the ESRF-EBS, all beam losses that occur can be explained by clearly identified causes and are fully different in appearance 
from the sudden loss phenomenon seen at KEK. 
There are also no reports, e.g., from the ESRF photon science users, of any sudden vertical blow up, with or without accompanying 
beam loss \cite{Qing}.}.  
For FCC-ee, neither MO-type vacuum flanges nor VacSeal \cite{baglin25} nor 
clearing electrodes will be used.  
Therefore, sudden beam losses in the form 
experienced at SuperKEKB are not expected to occur.

\section{Stored Beam Emittance}
\label{sec:sbe}
The design values for the rms vertical emittance are 8.6~pm in the LER
and 12.9~pm for the HER, as presented in Table \ref{tab:par}. 
These values correspond to a vertical-to-horizontal 
emittance ratio of 0.3\,\% for either beam.
The low-current emittances sometimes come close to this value, but often are larger.
The vertical emittance tuning and the achievable vertical 
emittance are ultimately limited by
the available beam diagnostics (Appendix \ref{sec:diag})
and by the vertical deformation of the 
SuperKEKB tunnel floor, which worsens every year (Appendix \ref{sec:tunnelfloor}).  

\begin{figure}[!ht]%
\centering
\includegraphics[width=0.80\textwidth]{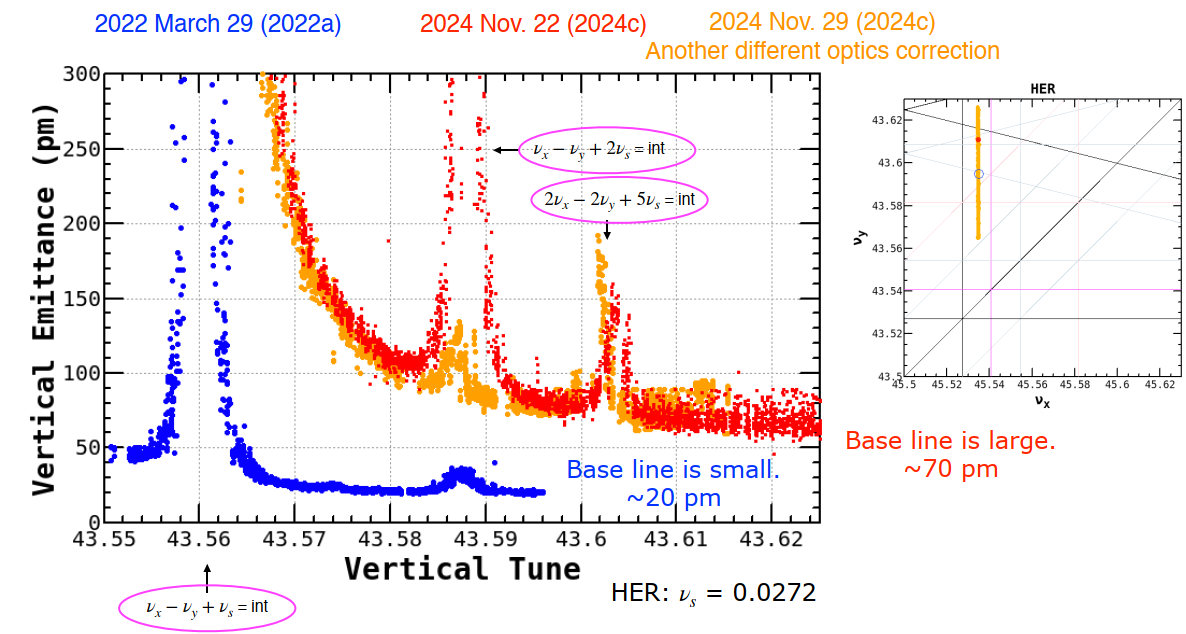}
\caption{Vertical emittance in the HER as a function 
of the vertical betatron 
tune $Q_y$ in November 2024 after two successive optics
corrections (red and orange curves) compared with 2022 (blue curve), 
as reported by Y.~Ohnishi \cite{ohnishi25}. 
The right picture shows the trace of this tune scan 
in the $Q_y$-versus-$Q_x$ tune diagram (red) along with
low-order resonances (black lines).  
}\label{fig:emit2024}
\end{figure}

During the early commissioning of the 2024c run, 
after two weeks of vacuum conditioning, 
the minimum vertical emittance in the HER 
increased from 20 pm to a value above 50--100~pm, 
and it did not decrease below 45--50~pm afterwards, even without collision 
and at low current.  
At the same time strong high order synchro-betatron resonances were observed 
in the tune scans without collisions, as is shown in Fig.~\ref{fig:emit2024}. 
Optics corrections could not recover the small vertical emittance, not even with a ``relaxed'' optics. 
The reason for the larger HER 
vertical emittance in 2024 is not yet fully understood. 
It suddenly appeared, perhaps as a result of a heat-induced 
movement of a beam pipe with beam-position monitors (BPMs) 
and/or magnets (see below), and it was unrelated to the presence of collisions 
and independent of $\beta_y^{\ast}$. 
Attempts were made to reproduce the measured optics in simulations, by misaligning sextupole magnets and rotating quadrupole magnets in the model, but it was not easily possible to obtain a vertical emittance as large as 50~pm for any model 
optics resembling the measured beam optics \cite{sugimoto25}.

\begin{figure}[!ht]%
\centering
\includegraphics[width=0.57\textwidth]{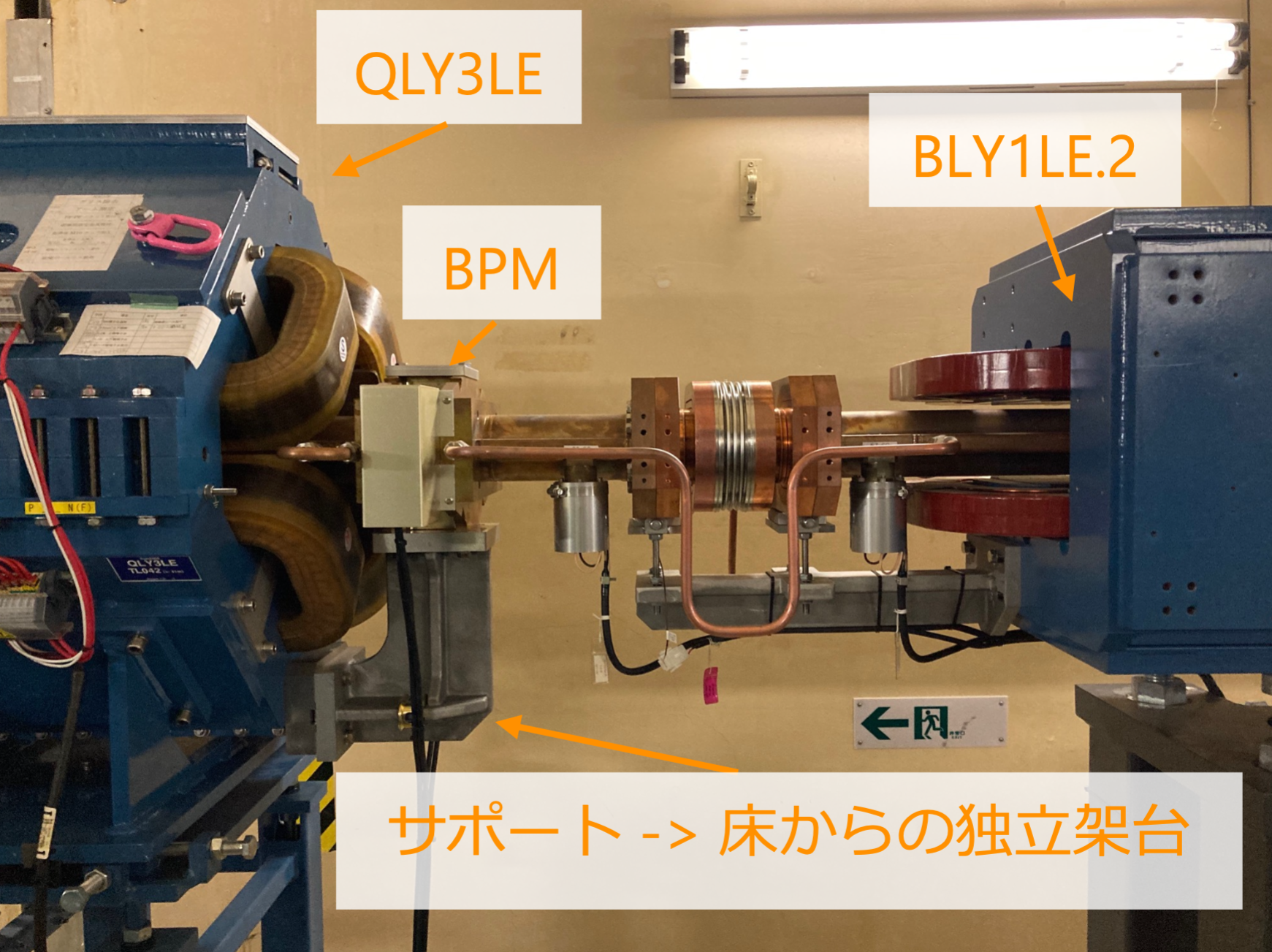}
\includegraphics[width=0.32\textwidth]{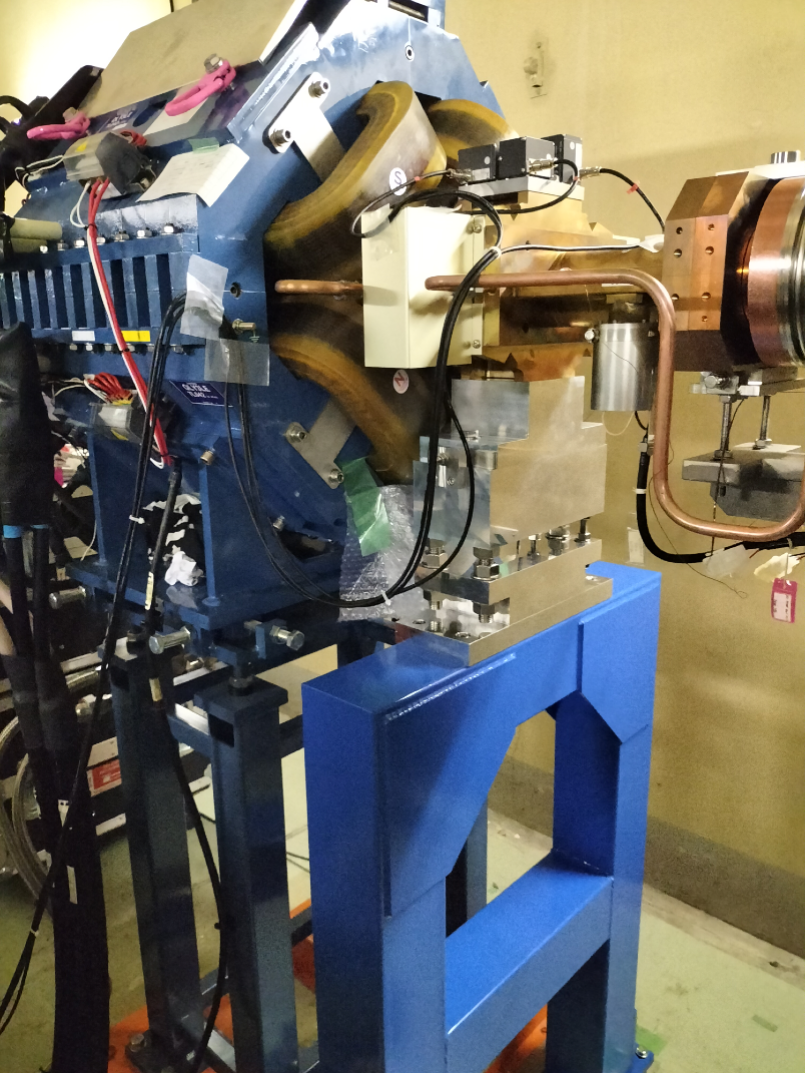}
\caption{Final focus BPM supported from the nearby quadrupole magnets prior to the 2024c run (left) 
and the BPM placed on its own support with vacuum 
chamber no longer connected to the magnet (right),  
as reported by Y.~Ohnishi \cite{ohnishi25}.}\label{fig:BPMsup}
\end{figure}

One hypothesis (mentioned by K.~Oide) is that the sudden HER emittance
growth could be related to a change of the mechanical support of critical BPMs in the final focus prior to the 2024c run. The beam pipe with BPM located between the strong sextupole pair for the local chromaticity correction had been supported by the nearby quadrupole magnet.
Deformation of the beam pipe due to synchrotron-radiation heating moved the beam pipe and pushed the quadrupole magnet away from its original position.
To avoid such a quadrupole motion, prior to 
the 2024c run, in one section of each ring, the beam pipe was decoupled 
from the quadrupole magnet and the BPM obtained its own support from the floor; as is illustrated in Fig.~\ref{fig:BPMsup}.
Now the synchrotron radiation heat no longer 
affected the position of the quadrupole magnet,
but it still moved the beam pipe with BPM, thereby  
losing the correlation between BPM readings and the beam orbit with respect to the adjacent magnets.

The vertical LER emittance
also increased by 
a factor of 2 from the 2024ab to the 2024c run.

\begin{figure}[!ht]%
\centering
\includegraphics[width=0.7\textwidth]{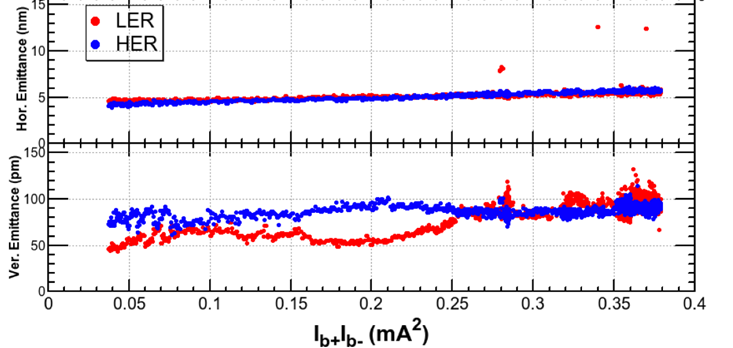}
\caption{Horizontal (top) and vertical emittance evolution (bottom)
for SuperKEKB HER and LER, observed in collision 
after reducing $\beta_{x}^{\ast}$ to 60~mm, 
as a function of the bunch current product, 
as reported by Y.~Ohnishi \cite{ohnishi25}.}
\label{fig:emitcol}
\end{figure}

At higher currents and in collision, 
the measured vertical emittances further increase to about 100 pm, and hence are about 10 times larger than the design. 
As an example, the 
vertical emittance evolutions with bunch current product in collision, as observed during  
the 2024c run, are displayed in Fig.~\ref{fig:emitcol}.

The FCC-ee design assumes a vertical-to-horizontal emittance ratio without collision of order 0.1--0.2\,\%.
Simulations with realistic optics errors indicates that this is achievable \cite{tomas:ipac2025-mopm009}.
Emittance ratios down to 0.01\%  
were demonstrated at some light-source storage rings, for example, 
at the Australian Synchrotron \cite{PhysRevSTAB.14.012804}. 
Based on simulations, the FCC-ee design also includes a 
50--100\,\% vertical emittance blow up in collision, which is consistent with the relative  blow up experienced at past and present colliders, 
including at SuperKEKB (e.g., Fig.~\ref{fig:emitcol}). 

At FCC-ee, the magnet supports will be sufficiently strong that 
they will not move when the nearby 
beam pipe is heated by synchrotron radiation.
In addition, the temperature of the FCC-ee beam pipe 
will be controlled by water cooling of 
dedicated photon absorbers. 
BPMs will be mounted on the same girders as the adjacent
sextupole and quadrupole magnets, and relative motion of these elements 
with respect to each other will be avoided, 
by installing any bellows only outside of such 
magnet-BPM combinations.

\section{Injected Beam Emittance}
\label{sec:ibe}
\begin{figure}[!ht]%
\centering 
\includegraphics[width=0.90\textwidth]{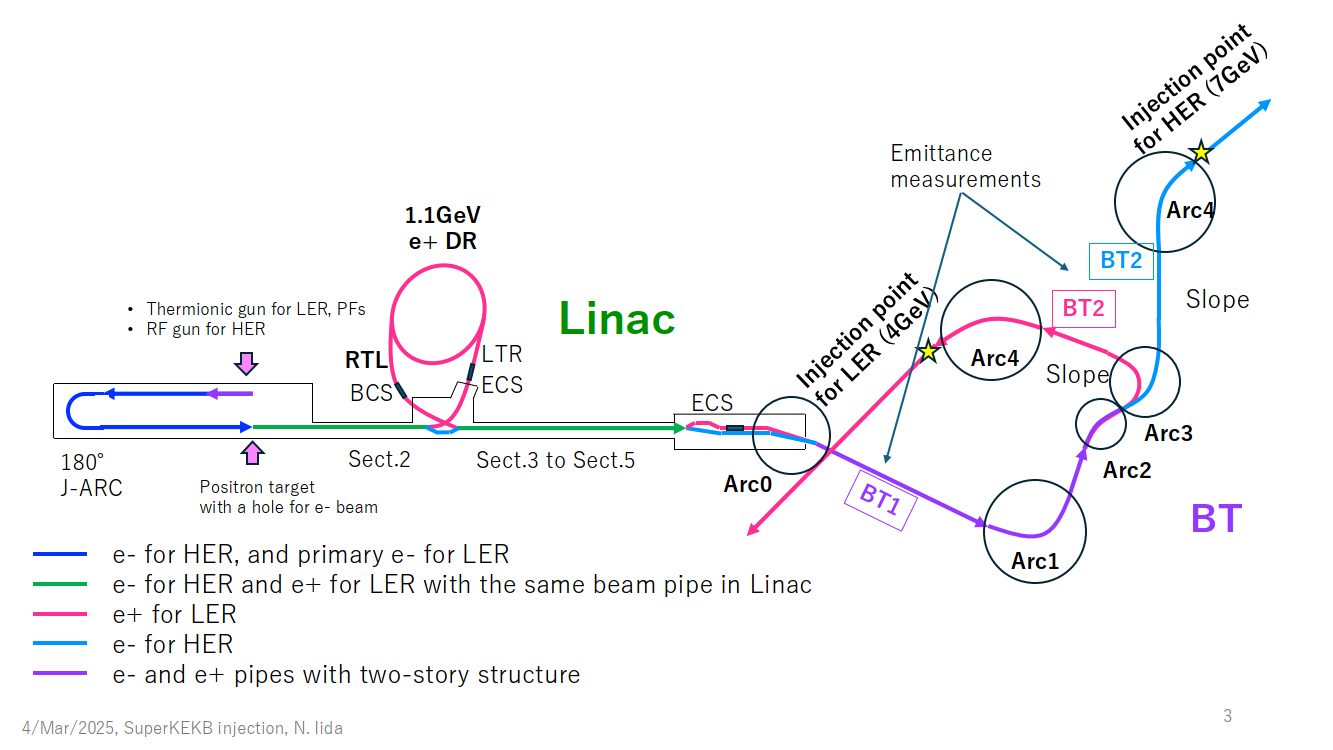}
\caption{Layout of the SuperKEKB injector and the beam transport (BT) lines (BTLs),
as presented by N.~Iida \cite{iida25b}. Each BTL comprises five arcs, which 
can be a source of emittance growth.}\label{fig:iida}
\end{figure}

The injector complex is illustrated in Fig.~\ref{fig:iida}.
From the end of the linac the electrons and positrons are sent through their respective 
beam transport lines BT1 and BT2. 
These lines and their magnets date back to the days of TRISTAN operation in the 1980s and 90s, at which time the 
beam emittances were orders of magnitude larger. 
Quite some bending is involved, especially for the positrons.
Emittances are measured at the locations marked ``BT1'' and ``BT2''. 
Large emittance growth is observed between these two locations for both beams and in both planes, as is illustrated in Table \ref{tab:bts}.

\begin{table}[htbp]
\caption{Emittance growth in the beam transport lines
of the SuperKEKB injector measured in 2024, 
as reported by N.~Iida \cite{iida25}.
}
\label{tab:bts} 
\begin{center}
\begin{tabular}{l|cc||l|cc} 
 \hline\hline
e$^+$ & BT1 & BT2 & e$^{-}$ & BT1 & BT2 \\
\hline
$\gamma \varepsilon_{x}$ & $ 110\pm 10$~$\mu$m &
$169\pm 16$~$\mu$m &  
$\gamma \varepsilon_{x}$ &  $39.2\pm 9.6$~$\mu$m & $142\pm 37$~$\mu$m \\
$\gamma \varepsilon_{y}$ & $10.5\pm 4.0$~$\mu$m & $81 \pm 20 $~$\mu$m &  
$\gamma \varepsilon_{y}$ & $42.5\pm 7.7$~$\mu$m &  $136\pm 22$~$\mu$m \\
\hline\hline
 \end{tabular}
 \end{center}
 \end{table}
 
\begin{figure}[!ht]%
\centering
\includegraphics[width=0.52\textwidth]{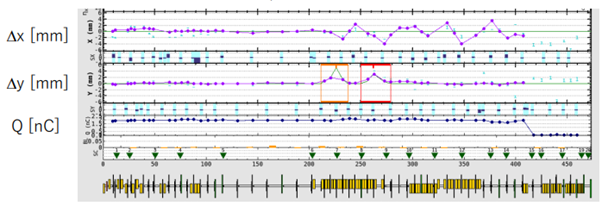}
\includegraphics[width=0.45\textwidth]{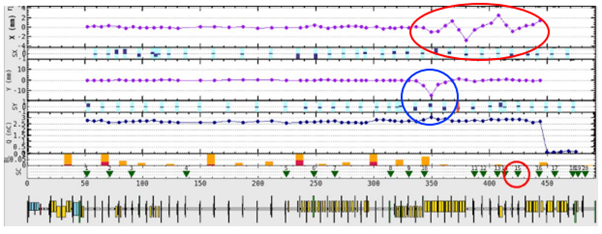}
\caption{A local vertical orbit bump generated a large horizontal orbit oscillation of comparable amplitude,
in both SuperKEKB HER BT (left) and LER BT (right), as reported by 
N.~Iida\cite{iida25}.}\label{fig:BTcoupling}
\end{figure}

Presently, the horizontal emittance 
growth for the electron beam (HER) is attributed to (in)coherent 
synchrotron radiation, which can explain 
an increase of the normalised horizontal 
emittance by about 75~$\mu$m \cite{iida25b}. 
For the LER, at the end of November 2024  
a correction of the horizontal dispersion 
in the damping-ring extraction line
reduced the horizontal emittance at BT1,
but a noticeable growth remained between BT1 and BT2 \cite{iida25b}. 
Regarding the vertical blow-up, unexpected nonlinear 
field errors were found to exist 
in the BT dipole magnets for either beam. 
In addition, strong local sources 
of horizontal-to-vertical coupling were detected in both 
electron and positron transport lines \cite{iida25},
as is illustrated in Fig.~\ref{fig:BTcoupling}.

In presence of the large blow up, the vertical emittance of the injected beam is about a factor 1000 larger than the design emittance of the stored beam. 
The enormous vertical emittances of the injected electron and positrons beams degrade the injection efficiency and
hinder a further squeeze of  $\beta_{y}^{\ast}$, since 
 in the interaction region (IR)
 the normalised physical aperture
decreases when reducing  $\beta_{y}^{\ast}$, 
becoming too small for the large vertical emittance of the 
injected beam, as discussed in 
Appendix \ref{sec:irap}. 

For FCC-ee, the injected  beam will not arrive directly from a linac, but from a full-energy booster with radiation damping and its own equilibrium emittance, and 
the FCC-ee transfer lines will not feature any sharp or large bends.
Based on the experience with 
other transfer lines at CERN, 
e.g., those built for LEP or the LHC,  
we also do not anticipate any large
local sources of betatron coupling.  
The emittances of the beam injected into the FCC-ee collider rings are expected to be similar to, or, possibly even smaller than, the stored beam emittances. 
The normalised IR aperture is at least an order of magnitude larger than for SuperKEKB, even when  
including a factor 10 margin for the injected beam emittance. 
Therefore, at FCC-ee, 
the vertical emittance of the injected beam 
(which could be up to 10\,\% of the booster horizontal emittance)  
should not limit the vertical 
interaction-point (IP) beta function $\beta_y^{\ast}$, and neither 
the injection efficiency, assuming the simulated 
values for the FCC-ee dynamic aperture 
and momentum acceptance, and the design values for the 
physical IR aperture.

\section{Injection Efficiency and Beam Current}
\label{sec:ie} 
The SuperKEKB 
injection is set up with two bunches per pulse at 25 Hz.   
In 2024,
charges of about 2.3 nC or 0.69 nC per pulse could be injected, respectively, into the LER and the HER.
The injection efficiency measured 100 turns after injection was about 80\% for the LER and 60\% for the HER.
The interaction-region (IR) 
aperture bottlenecks and associated collimator settings 
at $\beta_y^{\ast}=1$~mm amount to a 
normalised vertical aperture smaller than 3$\sigma$, 
if expressed in terms of the injected beam size (Appendix \ref{sec:irap}).  
With an expected beam lifetime of 
about 5 (LER) or 17 minutes 
(HER), respectively, reaching the intermediate target luminosity of $10^{35}$~cm$^{-2}$s$^{-1}$ would require
 a doubling of the effective injected positron 
 bunch charge \cite{iida25}.
The SuperKEKB record luminosity at the end of 2024 
was achieved in conditions where the 
injection was approximately saturated and the 
beam current could not be increased.

The SuperKEKB design beam currents are 
a factor 2, or more, larger 
than the maximum beam currents that 
could be achieved so far.
The beam currents are limited by the 
injection efficiency  
and bunch charges available from the linac.
However, the achieved beam current is already higher than the highest beam current values considered in the FCC-ee design, for operation on the Z pole. 
The FCC-ee injector linac will operate with 4 bunches per pulse, of up to about 5 nC each, at 100 Hz, and, therefore, with more than an order of magnitude higher injection rate.  
For FCC-ee, the collimators are set at 50~$\sigma_y$ of the stored beams,
corresponding to about 70~$\sigma_y$ of the expected injected beam,
and still more than 20~$\sigma_y$ with a factor 10 margin for 
possible emittance blow up in the booster or during transfer.

\section{Interaction Region}
\label{sec:IR} 
\begin{figure}[!ht]%
\centering
\includegraphics[width=0.90\textwidth]{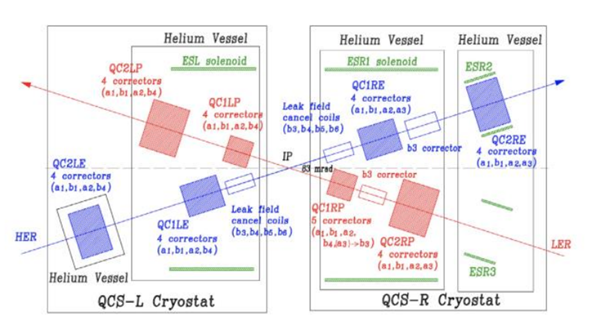}
\caption{Layout of the superconducting IR magnet system of SuperKEKB \cite{masuzawa24}.}
\label{fig:qcs}
\end{figure}

The interaction region (IR) of SuperKEKB
is complex. A discussion of the 
IR aperture bottlenecks is deferred to 
Appendix \ref{sec:irap}.  
Figure \ref{fig:qcs} shows the 
conceptual layout of the final focusing 
superconducting magnets QCS. 
The QC1 and QC2 magnets provide  vertically and horizontally 
focusing fields, respectively. 
The QCS magnet systems comprise eight individual 
quadrupole magnets, 
thirty-five corrector magnets for beam tuning, 
eight magnets for leak field cancellation, 
and four solenoids, called 
ESL, ESR1, ESR2, and ESR3.
The effect of the solenoid field of Belle II is fully compensated by the compensation solenoids 
ESL, ESR1, ESR2, and ESR3, if the condition 
$\int B_{z}(s) \, ds =0$ is met.
Since at SuperKEKB the compensation solenoids and quadrupole fields are superimposed, and, in addition, the solenoids 
are arranged in a non-symmetric way, it is unlikely that this cancellation condition is exactly fulfilled. 
An imperfect cancellation would give rise to a residual 
chromatic skew coupling at the collision point,
which is not easily measured and corrected.
Determining the exact magnetic fields encountered by the beams 
on their complicated trajectory through this region,
and correctly modeling the resulting higher-order
beam optics is challenging and necessarily prone to errors, 
as will be illustrated in the following.   

\begin{figure}[!ht]%
\centering
\includegraphics[width=0.80\textwidth]{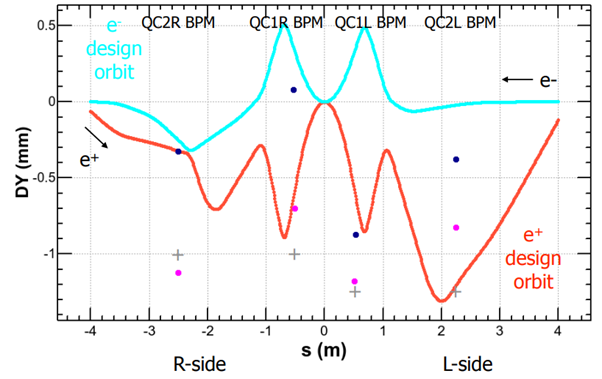}
\caption{Vertical design orbits of electron (light blue curve, from right to left) and positron beams (red curve, from left to right) around the SuperKEKB interaction point and the measured beam positions (dots), 
shown by Y.~Ohnishi 
\cite{ohnishi25}. 
The dark blue points represent the HER beam orbit measured at the QC1LE and QC2LE BPMs. 
The pink points correspond to the
measured LER orbit.
The ``+'' signs indicate the vertical offset from a straight reference orbit of the LER BPMs 
at QC1LP, QC2LP, QC1RP and QC2RP.  
The magnets have the same (or a similar) mechanical offset as their associated BPMs. 
The BPM readings have been  corrected for 
the mechanical offsets. 
}\label{fig:ir}
\end{figure}

Figure \ref{fig:ir} presents the 
design beam orbits of electrons and positrons 
in the vertical plane (the solid curves).  
The measured beam orbits significantly deviate from the 
model prediction. 
The beam measurements (dots) indicate an 
average vertical offset between electrons and positrons of about 400 $\mu$m along with a 400 $\mu$rad vertical angle between the two colliding beams, whereas the design collision scheme has zero offset in vertical position and angle.

In addition to the unexplained large orbit errors
around the interaction region, also an unexpected shining of 
synchrotron radiation onto the inner detector was observed. 
A peculiar steering of the orbit is required to bring the beams in collision and to avoid this shining. 
A speculation is that the alignment of the Belle-II detector 
might have large errors, in position and angle, relative to 
the nearby accelerator components \cite{oide25}.
In simulations, a 1 mrad vertical tilt angle of the solenoid at the IP 
results in a vertical emittance of about 70 pm in the HER,
which after coupling and dispersion correction decreases to
16 pm \cite{oide25}. 
The vertical emittance before and after correction increases roughly with the square of the detector tilt angle.

\begin{figure}[!ht]%
\centering
\includegraphics[width=0.8\textwidth]{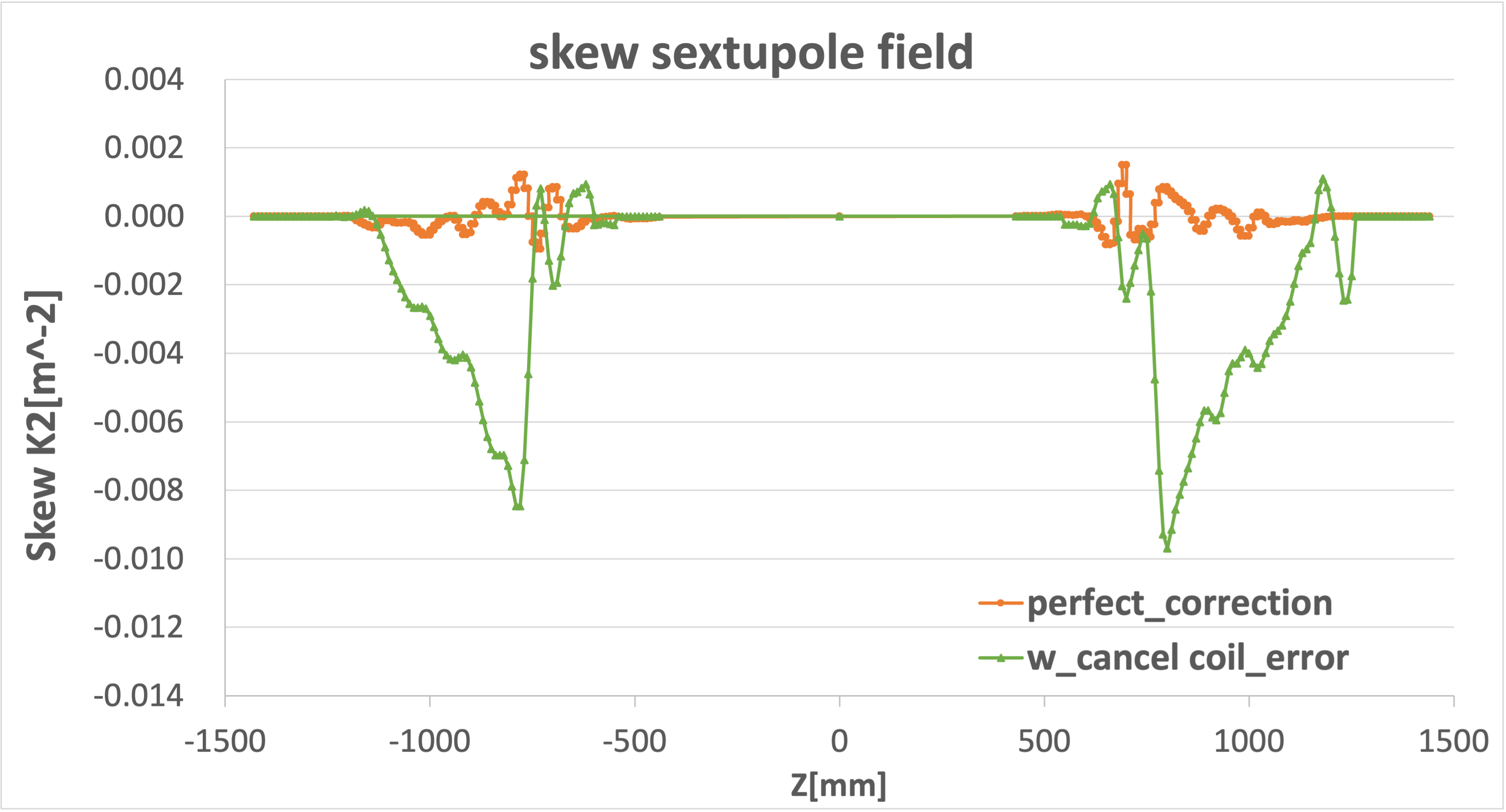}
\includegraphics[width=0.8\textwidth]{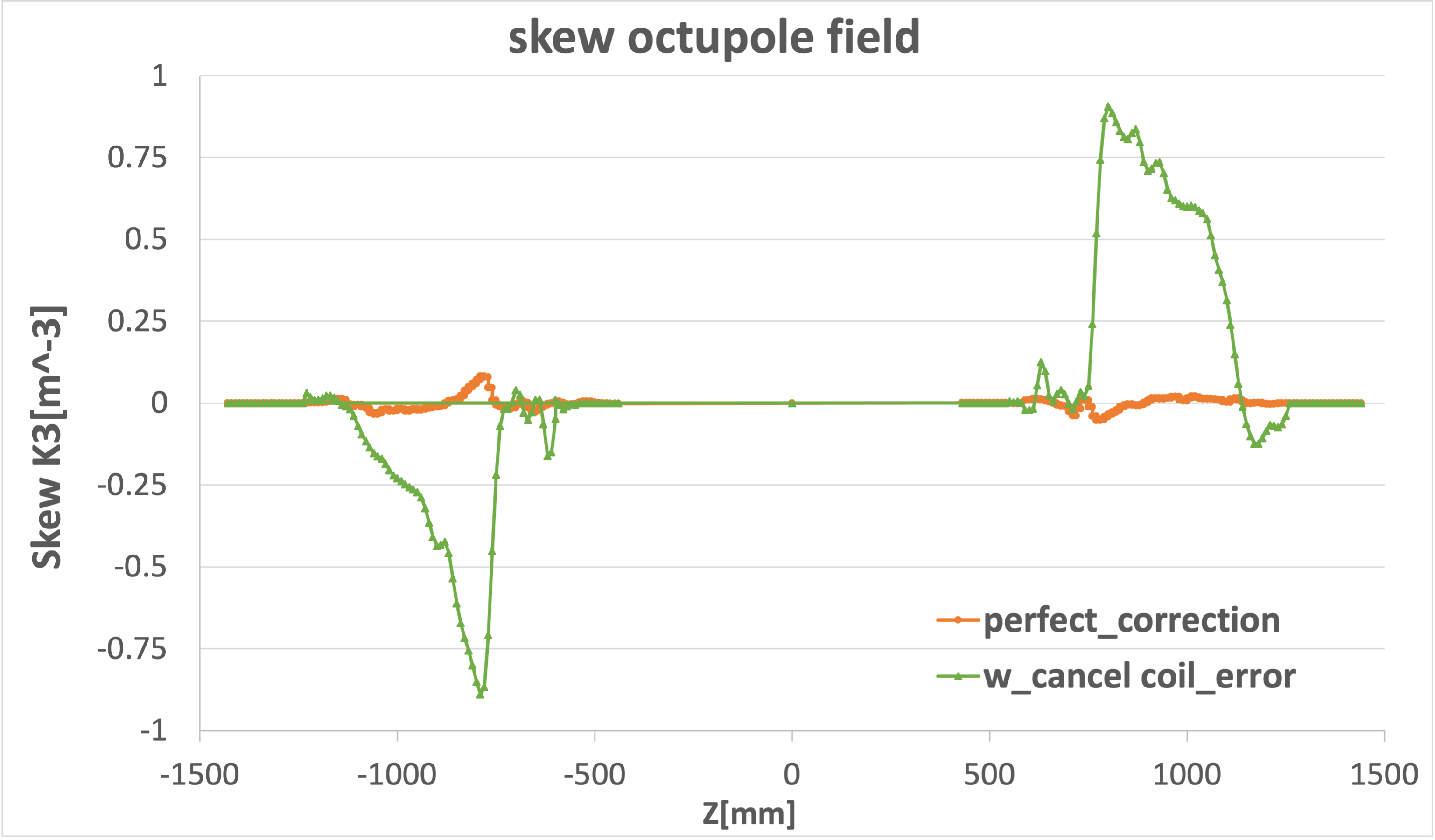}
\caption{The design (orange) and actual (green) leakage skew sextupole and octupole field components, generated by the LER final quadrupoles, as seen by the HER beam, versus the distance from the collision point, shown by L.~Meng \cite{limeng} based on input from Y.~Ohnishi.}\label{fig:ircerr}
\end{figure}

Aggravating the situation is a magnet coil construction error.
Cancel coils in the HER should correct the leakage field 
originating from the final quadrupoles QC1LP and QC1RP of the LER,
which affect the HER electron beam.   
Namely, all skew sextupole ($a_3$) and skew octupole ($a_4$) cancel coils are being powered in series together with the 
normal sextupole ($b_3$) and octupole ($b_4$) corrections, 
but the polarity of the $a_3/_4$ coils was mistaken during construction, so that when correcting the $b_3/b_4$ errors, the $a_3/a_4$ errors are actually doubled, instead of being corrected.
The situation is illustrated in Fig.~\ref{fig:ircerr}.

A recent simulation by M.~Li studied the consequences \cite{limeng}. 
These tracking studies reveal that the cancel coil
errors degrade the on-momentum dynamic aperture from 80$\sigma_y$
to 65$\sigma_y$, and from 22$\sigma_x$ to 16$\sigma_{x}$
(with geometric rms emittances of $\varepsilon_{x}=4.5$~nm
and $\varepsilon_y=25$~pm, which is up to
four times smaller than the typical vertical 
emittance in collision),  
and decrease the injection efficiency by more than 10\% (with injected beam emittances of 
$\varepsilon_{x}=25$~nm and $\varepsilon_y=20$~nm).

The two colliding FCC-ee beams have (approximately) the same energy (with a most a few percent energy 
difference for the ${\rm t}\bar{\rm t}$ running),   
and a simpler interaction region is envisaged. 
The FCC-ee interaction region will include 
a full local compensation of coupling and chromatic coupling 
from the detector solenoid.  
Instead of staggered magnets, alternating for the two beams,
it will feature final quadrupoles with two apertures focusing both  beams.  By virtue of the canted-cosine-theta 
magnet design concept, the cross-talk between the  
(nonlinear) magnetic fields of the two apertures is zero, 
and higher-order field errors 
are expected to be insignificant \cite{8292834}. 
Also, at FCC-ee, 
compensation and shielding solenoids in the interaction region 
locally cancel the effects of the detector solenoid on the beam 
\cite{koratzinos2021magneticcompensationschemefccee}, 
 with negligible overlapping of quadrupole and solenoid fields.  

\section{Vertical Beta Function at the Collision Point}
\label{sec:vb}
\begin{figure}[!ht]%
\centering
\includegraphics[width=0.75\textwidth]{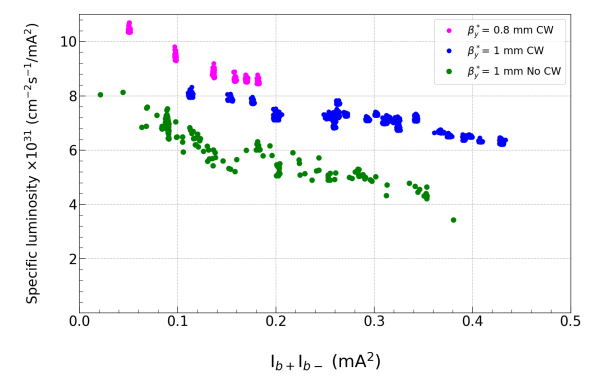}
\caption{Measured specific luminosity as a function of bunch-current product, for $\beta_y^{\ast}$ values of 0.8 and 1.0~mm, including crab waist, and 
at $\beta_y^{\ast}=1.0$~mm
also without, 
presented by Y.~Ohnishi in 2023  \cite{ohnishi23}.}
\label{fig:beta08}
\end{figure}

In  Run  2021c--2022b, the vertical beta functions were reduced to
$\beta_y^{\ast}= 0.8$~mm, leading to a significantly higher specific luminosity 
at low bunch currents.   
The specific luminosity for $\beta_{y}^{\ast}$= 0.8~mm was 
about 20\,\% higher than for 1~mm at a bunch-current product of 0.1~mA$^2$.
The gain in specific luminosity 
decreased to about 12\,\% at 0.16~mA$^2$ and above.
These observations  are illustrated in Fig.~\ref{fig:beta08}. 
The design 
bunch current product is about 1.5 mA$^{2}$.  

The injection efficiency and the beam lifetime also degrade with 
smaller $\beta_{y}^{\ast}$, while the detector background worsens.
In 2024, the vertical beta function
was not squeezed below a value of 1 mm.

FCC-ee does not require any vertical 
beta functions much smaller than 0.8--1.0\,mm, which were demonstrated at SuperKEKB ---
with about half the effective focal length and 
about 100 (10,000) times larger vertical emittances of the stored (injected) beams than at FCC-ee.

\section{Specific Luminosity}
\label{sec:specl} 
\begin{figure}[!ht]%
\centering
\includegraphics[width=0.75\textwidth]{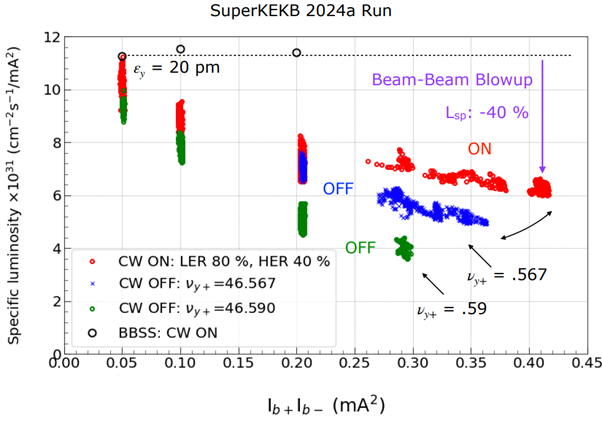}
\caption{Measured specific luminosity as a function of bunch-current product without the crab waist at two betatron tunes (green and blue) and with the nominal crab waist (red), 
presented by Y.~Ohnishi  \cite{ohnishi25}.
 }\label{fig:cw}
\end{figure}

At higher bunch-current products the SuperKEKB specific luminosity decreases.  
The crab waist collisions scheme improves the situation, 
as is shown in Figs.~\ref{fig:beta08} and \ref{fig:cw}, but it does not fully  restore the value 
achieved at very low bunch currents.
  
A significant reduction of the SuperKEKB specific luminosity at higher bunch  
currents had long been expected from simulations  \cite{Zhou:2015cva}. 
This predicted reduction has never fully been included in the advertised  design luminosity. 

\begin{figure}[!ht]%
\centering
\includegraphics[width=0.75\textwidth]{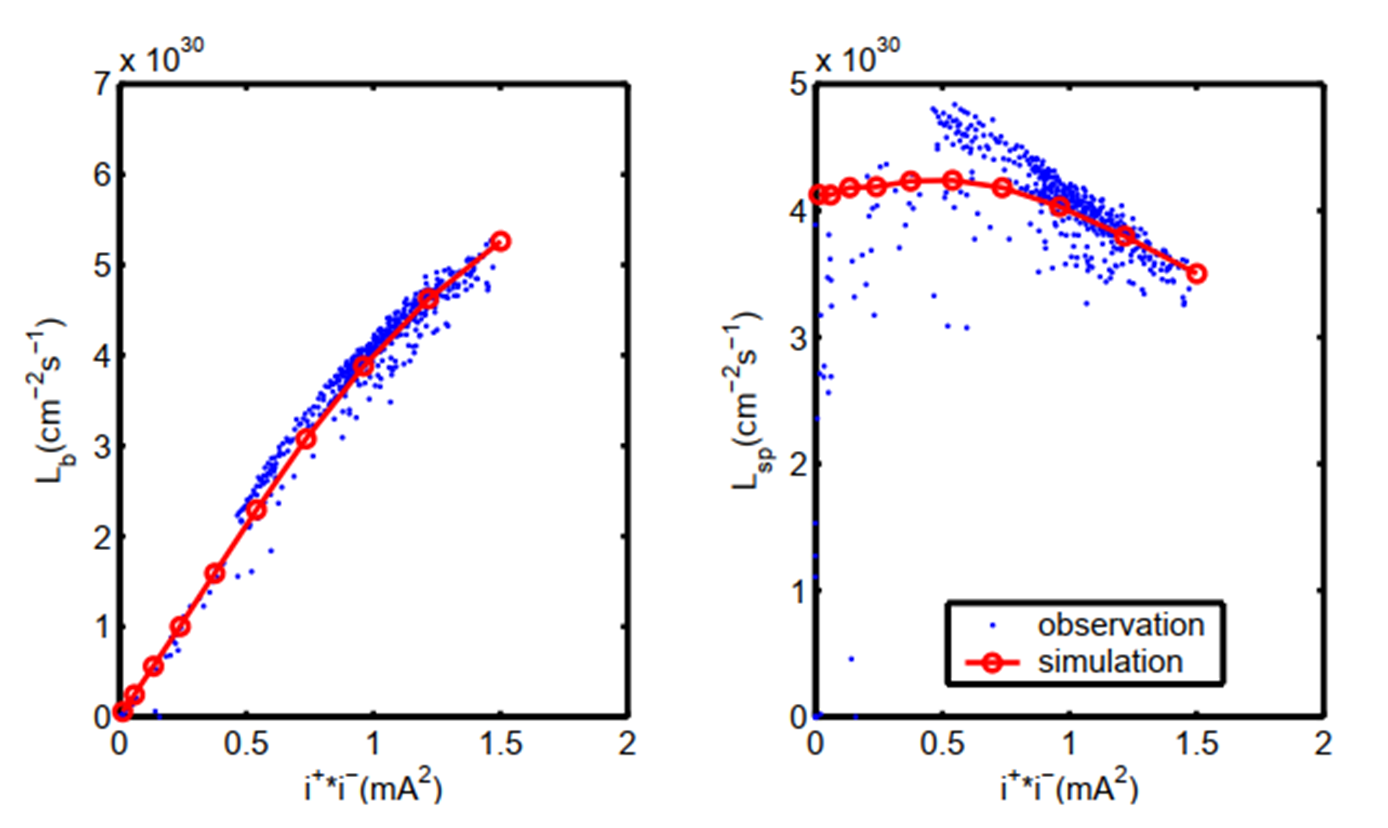}
\caption{Measured (blue) and simulated (red) luminosity (left) and specific luminosity 
(right) as a function of the product of beam currents for head-on collisions at PEP-II, 
presented by Y.~Cai in 2005 \cite{yunhai05}. 
The simulation included the full lattice nonlinearity. }
\label{fig:cai}
\end{figure}

The importance of adding the nonlinearity of the lattice
to the beam-beam simulations had earlier been demonstrated for the former 
SLAC PEP-II B factory, where beams collided with zero crossing angle \cite{yunhai05}. 
When including  the lattice nonlinearities, for PEP-II the agreement between simulated and measured 
specific luminosity was at the level of 10\%, as is illustrated in Fig.~\ref{fig:cai}.  
Even the computed beam lifetimes agreed well 
with the observations \cite{yunhai05}.

For SuperKEKB, both the effects of the full nonlinear lattice and 
the space-charge forces are expected to be important. 
In the LER, the space-charge effect can produce a tune spread of almost the same value and with opposite sign to that of the beam-beam kick. Since the dependence of the space charge tune shift on the longitudinal position $z$ along the bunch is opposite to the 
variation of the beam-beam tune shift with $z$, these two effects can conspire so as to reduce the achievable beam-beam tune shift and to degrade the specific luminosity. 

\begin{figure}[!ht]%
\centering
\includegraphics[width=0.95\textwidth]{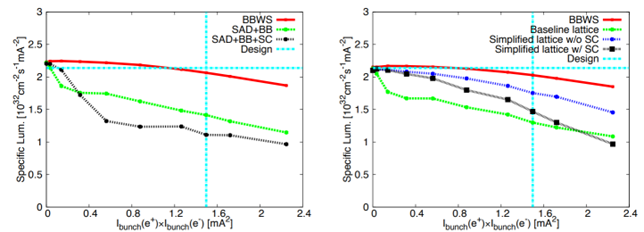}
\caption{Specific luminosity as a function of bunch current
product obtained from 
weak-strong simulations for the LER, considering an ideal situation with a linear machine (``BBWS'', red curve), 
a realistic baseline lattice with detector solenoid 
(``SAD'', green curve), the effect of space charge (SC) added to a simplified lattice (black curve, right picture), or the combined effect of the full lattice and space charge 
(black curve, left picture),  presented by D.~Zhou in 2015 \cite{Zhou:2015cva}. 
The black curve with space charge in the right picture can be compared with the blue curve presenting the effect of the simplified lattice only; the latter contains simpler final-focus quadrupoles and no solenoids 
 \cite{Zhou:2015cva}.}
\label{fig:zhou15a}
\end{figure}

\begin{figure}[!ht]%
\centering
\includegraphics[width=0.85\textwidth]{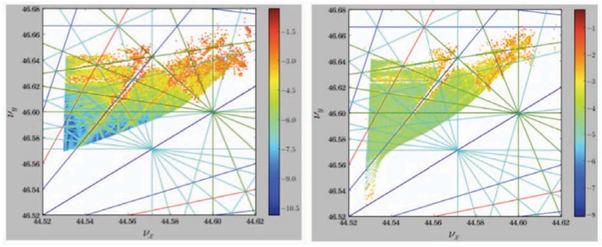}
\caption{Simulated betatron tune footprint for a lattice of the SuperKEKB LER, including the beam-beam collision,  without (left) and with (right) space-charge effect, 
presented by D.~Zhou in 2015  \cite{Zhou:2015cva}. 
Resonance lines up to 7th order are indicated.}
\label{fig:zhou15}
\end{figure}

The simulated dramatic effects of a realistic lattice and of 
LER space charge forces on the specific luminosity,
as reported at IPAC 2015 \cite{Zhou:2015cva},  
are illustrated in Figs.~\ref{fig:zhou15a} and
~\ref{fig:zhou15}. 
Noteworthily, most of the
simulations in Fig.~\ref{fig:zhou15a} did not  
achieve the design luminosity value of 
$8\times 10^{35}$~cm$^{-2}$s$^{-1}$,
corresponding to a nominal specific luminosity of  
 $2.14\times 10^{32}$~cm$^{-2}$s$^{-1}$mA$^{-2}$.  
Only the ideal weak-strong beam-beam simulation 
without any nonlinear accelerator lattice and not including 
space charge came close to it.  
This same finding also remains true for the  running conditions in 2024:   
Recent simulations, using the APES code \cite{Zli},  
considering actual SuperKEKB parameters, with 0.7 mA bunch current in the LER
and 0.54 mA in the HER (hence, for a bunch-current product of 0.38~mA$^2$, 
still far below the design value of 1.5~mA$^2$)  
show a 32\,\% drop in specific luminosity when the nonlinear lattices of LER and HER  
are added to the beam-beam simulation \cite{zhiyuan}.

We also note that for the SuperKEKB parameters of 2022 or 2024, the LER space charge tune shift was about
$\Delta Q_{y,{\rm SC}}\approx -0.02$, and hence within a factor of two equal to the
magnitude of the beam-beam tune shift in those runs \cite{jack}.

\begin{figure}[!ht]%
\centering
\includegraphics[width=0.90\textwidth]{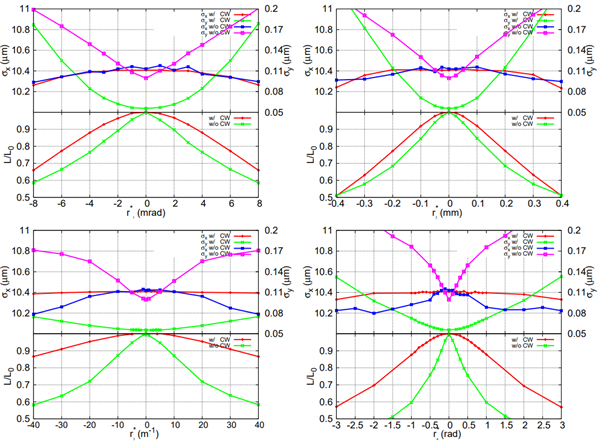}
\caption{Simulated beam sizes and relative luminosity as a function of
the linear $x$-$y$ coupling parameters at the IP with and without the 
crab waist collision scheme, presented by D.~Zhou in 2010, 
where he considered slightly different  
SuperKEKB design parameters \cite{Zhou:2010zzd}.}
\label{fig:IR-C}
\end{figure}

\begin{figure}[!ht]%
\centering
\includegraphics[width=0.75\textwidth]{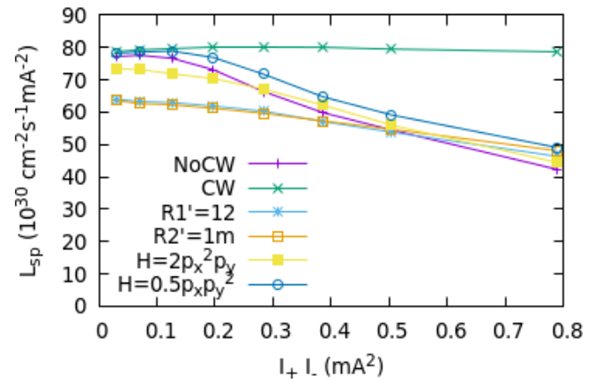}
\caption{Simulated specific luminosity  as a function of bunch-current product,
for different values of the chromatic coupling $R_1'=dR_1^{\ast}/d\delta$,
and $R_2'=dR_2^{\ast}/d\delta$,  presented by K.~Ohmi  \cite{ohmi25}.}
\label{fig:IR-CC}
\end{figure}

These and additional studies have demonstrated 
that the specific luminosity of SuperKEKB 
is reduced by any of the following effects:
\begin{itemize}
\item significant space-charge forces in the LER;  
\item nonzero linear coupling at the collision point for the HER, since this coupling could not fully be corrected by the available IP coupling knob; in fact, the $x$-$y$ coupling knob at the X-Ray Monitor indicates the existence of ``dynamic coupling,'' caused by the beam-beam focusing along with significant residual coupling at the IP ($R_1^{\ast}\approx -3.78\times 10^{-3}$~rad, $R_2^{\ast}\approx 0.54$~mm) \cite{oide25}; 
the IP coupling tolerances  
resulting in 20\% loss of specific luminosity
were determined for an earlier version of SuperKEKB, 
with crab waist, as 
 $|R_1^{\ast}| \le 5.3\cdot 10^{-3}$~rad, and $|R_2^{\ast}| \le 0.18$~mm \cite{Zhou:2010zzd} 
 (see Fig.~\ref{fig:IR-C}); 
 as the experimentally inferred 
 value of $R_2^{\ast}$ is three times this tolerance, 
 a loss in specific luminosity,  
 due to residual linear IP coupling, 
 of more than 50\% should be expected;  
\item  significant nonzero chromatic coupling due to the complex 
antisolenoid scheme and due to the IR coil manufacturing error, which causes a large uncorrected $a_3$ field component;
the simulated effect is significant (Fig.~\ref{fig:IR-CC}); 
also other, higher-order aberrations, e.g., as due to the $a_4$ IR coil error, may play a role; 
\item noise excitation by the bunch-by-bunch feedback system and/or an interplay of the latter with the machine impedance \cite{PhysRevAccelBeams.26.071001}; with only a few colliding bunches the specific luminosity is higher when the feedback is turned off; see  Fig.~\ref{fig:noise};
\item bunch lengthening due to the longitudinal impedance (see Appendix \ref{sec:bd}).
\end{itemize}

Figures \ref{fig:IR-C} and \ref{fig:IR-CC}
present the simulated effects of linear coupling  and 
chromatic coupling at the IP, respectively,  
on the specific luminosity.
They are both similar to the experimentally observed decrease.
As already 
mentioned, the estimated residual linear coupling,  
inferred from beam measurements in 2024, was of order
$R_1^{\ast}\equiv r_1^{\ast}\approx -3.8$~mrad, and 
$R_2^{\ast}\equiv r_2^{\ast} \approx 0.54$~mm \cite{oide25}, 
which might be compared with the sensitivity displayed in the two top pictures of Fig.~\ref{fig:IR-C}, 
though the SuperKEKB optics and beam parameters considered in the year 
2010 were not exactly the same. 



The measured effects of the bunch-by-bunch feedback and of a simulated feedback noise on the specific luminosity is illustrated
in Fig.~\ref{fig:noise}.

\begin{figure}[!ht]%
\centering
\includegraphics[width=0.45\textwidth]{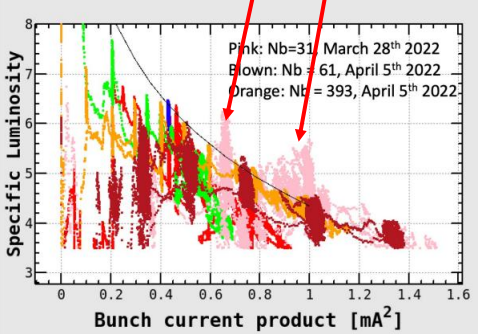}
\includegraphics[width=0.45\textwidth]{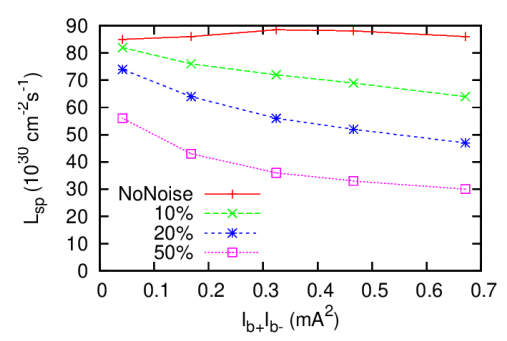}
\caption{Measured (left) and simulated specific luminosity (right) as a function of bunch-current product, presented by K.~Ohmi  \protect\cite{ohmi25}.
The pink data on the left were taken with transverse 
bunch-by-bunch feedback system switched off, while the feedback was active, with more bunches, for the other colors.
Different curves on the right consider various levels of feedback noise, expressed in percent of the rms vertical
beam size. }
\label{fig:noise}
\end{figure}

The situation for SuperKEKB is quite different from
the FCC-ee design, where the nominal luminosity was chosen based on the results of weak-strong beam-beam simulations  with the complete nonlinear 
accelerator optics, including interaction-region solenoids \cite{fsrv2}. 
Quasi-strong-strong beam-beam simulations with the full ring optics 
were also carried out to 
study the alternating injection of 
electron and positron bunches into FCC-ee \cite{mori25}.   
Several different simulation tools, developed at KEK, BINP, IHEP, and CERN/EPFL, confirmed the simulated FCC-ee design luminosity values.
The sensitivity of the luminosity and beam lifetime 
to optics errors and to emittance imbalances is also being 
determined, e.g., in Refs.~\cite{PhysRevAccelBeams.27.121001,leon2025}.
FCC-ee alignment and field-error 
tolerances are defined accordingly. 
In parallel, 
the impact of the machine impedance in FCC-ee performance 
and associated mitigation measures are under study \cite{yzhang22,soos2025}. 
The space charge forces in the FCC-ee collider rings are much 
smaller than for the SuperKEKB LER, thanks to the higher beam energy and longer bunches.


\section{Summary and Conclusions}
\label{sec:sc}
SuperKEKB delivers more than two times
higher luminosity than the previous KEKB, at 
much reduced synchrotron radiation power.
It routinely operates with the virtual crab waist collision
scheme first developed for FCC-ee, 
with vertical beta functions of 1\,mm (or below)  
as required for the FCC-ee Higgs factory running,  
and with beam currents higher than the 
maximum FCC-ee values foreseen on the Z pole.
Three key concepts of the FCC-ee design have, thereby, been validated.

In general, the SuperKEKB experience is providing important lessons and input for the FCC-ee design effort.   
The SuperKEKB luminosity performance is still an order of magnitude short of the target  value. As discussed in this report, a 
number of issues are limiting the present SuperKEKB luminosity performance.
All of these appear to be very specific to SuperKEKB.
They were not observed, 
in this form, at any previous colliders or at any of the modern light sources.
A few additional and complementary issues faced by SuperKEKB are addressed in the Appendix. 
We see no reason or argument to suggest 
that any of these problems should be encountered at FCC-ee.
Specifically, we highlight the following points.
\begin{enumerate}
\item For FCC-ee, no MO-type vacuum flanges will be used.
Also it is not planned to apply 
any VacSeal for tightening leaks at FCC-ee \cite{baglin25}.    
For both reasons, sudden beam loss events like 
those observed at SuperKEKB are not expected to occur.
\item The beam injected into the FCC-ee collider rings 
is delivered from a full-energy booster. which should provide a vertical beam emittance similar to (or even smaller than) the design emittance 
of the collider and not an emittance  several orders of magnitude higher.
Nevertheless a factor ten margin for the vertical emittance 
is included in the FCC-ee injection design. 
\item A vertical-to-horizontal emittance ratio without collision
of order 0.1--0.2\,\%, as assumed in the FCC-ee design, should be achievable, as indicated by simulations with realistic key 
optics errors \cite{tomas:ipac2025-mopm009}. 
Even much smaller emittance ratios, down to 0.01\%, 
were demonstrated at various light sources, e.g., at the Australian Synchrotron
\cite{PhysRevSTAB.14.012804}. 
\item The FCC-ee interaction region will be simpler, it will have an at least an order-of-magnitude larger aperture normalised to the
size of the injected beam, and it will include 
a full local compensation of coupling and chromatic coupling.  
\item 
The beam-position monitors at FCC-ee will be fixed on a common girder with the nearby quadrupole and sextupole magnets, and the magnet supports will be so robust and strong, that the magnets will not move sideways when the beam pipe is heated by synchrotron radiation.
\item  Most of the available beam diagnostics at SuperKEKB does not have turn-by-turn, bunch-by-bunch detection capability and is insufficient for measuring the beam optics in collision or analysing plus correcting nonlinear terms. FCC-ee will have 
 state-of-the-art optics diagnostics like the BPM system of the LHC.
\item The SuperKEKB tunnel is floating and suffers from significant vertical deformation, increasing every year,  
which degrades the vertical beam emittances.
The deeper FCC-ee tunnel floor should be stabler, 
with the LEP/LHC tunnel serving as an example.   
\item Space-charge effects at FCC-ee are smaller thanks to the higher beam energy and longer bunches, and, for colliding bunches in FCC-ee, the bunch lengthening with current due to impedance 
is much weaker thanks to the dominant effect of beamstrahlung. 
\end{enumerate} 
For FCC-ee, the design luminosity values are based on weak-strong 
beam-beam simulations including the full nonlinear lattice \cite{fsrv2}.  
The effects of optics errors on the luminosity are also being evaluated and tolerances defined accordingly, e.g.,~\cite{PhysRevAccelBeams.27.121001,leon2025}. 
In addition, the interplay of the beam-beam interaction and the machine 
impedance is being investigated \cite{yzhang22,soos2025}.    
Although neither MO-flanges nor VacSeal will be used, 
the sudden-beam-loss experience of SuperKEKB has motivated an ongoing
study of other possible beam-dust interactions for FCC-ee.
The acceptable noise frequency spectrum  and the corresponding feedback design specifications for 
FCC-ee will be determined through beam-beam simulations. 

Table \ref{tab:compfinal} compiles the various challenges encountered by SuperKEKB 
together with the estimated resulting loss in luminosity with respect to the design
luminosity, and a similar column with the corresponding expectations for FCC-ee.

The different factors affecting the specific luminosity are not truly multiplicative. 
If the specific 
luminosity is already degraded by some error, then the other effects become weaker.
For example, if the emittance is bigger, or the bunch longer, the space-charge force and its impact become smaller.   
Similarly, the dominant optics aberration at the IP will determine the beam-beam performance and, thereby, render the effect of the other aberrations comparatively less important. 
In other words, once the beam blows up due to one error, 
the additional errors have a lesser impact on the specific luminosity. 
From the various simulations and actual measurements, we 
extrapolate that the total reduction factor in specific luminosity 
will not be smaller than a factor of 0.3 (which would be attained, e.g., with 100 pm vertical emittances in collision instead of the design values around 10 pm).
Adopting this value, the total SuperKEKB luminosity 
loss factor becomes: $0.3$ (inj.~beam emittance \& IR aperture) $\times$  0.3 (reduced spec.~luminosity) $=0.09$,
translating into a total luminosity of 
$7\times 10^{34}$~cm$^{-2}$s$^{-1}$.

For FCC-ee, the simulations used to define the 
design luminosity already consider the full nonlinear lattice, while the effects of space charge and bunch lengthening are small or negligible.
A factor ten margin is included for the 
injected beam emittance. 
It is further assumed that the FCC-ee IR coils will have no significant 
manufacturing errors, that the FCC-ee 
feedback noise will meet specifications, and that the targeted stored beam emittance without 
collision can be reached.
A factor of two blow-up in the vertical emittance of the stored beam due to the collision is also taken into account in the design parameters.  
With these inputs and assumptions, none of the problems afflicting SuperKEKB is expected to reduce the 
FCC-ee luminosity below its design value.
Of course, the FCC-ee may well encounter some other, different challenges. 

\begin{table}[htbp]
\caption{Issues encountered by SuperKEKB 
together with the estimated resulting luminosity loss factor with respect to the design luminosity at nominal beam current, 
and the corresponding expectation for FCC-ee.
}
\label{tab:compfinal} 
\begin{center}
\begin{tabular}{l|cc|c}
\hline\hline
effect / problem   &  SuperKEKB & FCC-ee & comment \\
\hline
{\bf injected beam vert.~
emittance}  \&  & {\bf $\sim$0.3} & 1.0 &  both $\beta_y^{\ast}$ squeeze and injection \\
\hspace*{5 mm} 
{\bf limited IR aperture }&  &  &   efficiency (beam current) affected  \\
\hline
stored beam vert.~emittance & $\sim$0.5 & $\sim$1.0 &  reduction in spec.~luminosity \\
\hspace*{5 mm} due to tunnel deformation, 
 &  &  &  \\
\hspace*{5 mm} non-anchored BPMs, etc. &  &  &  \\
nonlinear lattice & 0.6 & 1.0 &  reduction in spec.~luminosity \\
bunch lengthening & 0.6 & $\sim$1.0 &  reduction in spec.~luminosity \\
residual linear \& chromatic  & 0.5 & 1.0 & reduction in spec.~luminosity \\
\hspace*{5 mm} coupling at the IP &  &  &  \\
IR coil errors & 0.5 & 1.0 &  reduction in spec.~luminosity \\
space charge & 0.6 & $\sim$1.0 &  reduction in spec.~luminosity \\
feedback noise & 0.8 & $\sim$1.0 &  reduction in spec.~luminosity \\
{\bf combined effect of last 7 items}  & {\bf $\sim$0.3 }(?)  & $\sim$1.0 & 
total spec.~luminosity reduction\\ 
\hline
{\bf effective total loss factor} & {\bf $\sim$0.09} & $\sim$1.0 &  \\
\hline\hline
 \end{tabular}
 \end{center}
 \end{table}

\begin{figure}[!ht]%
\centering
\includegraphics[width=0.75\textwidth]{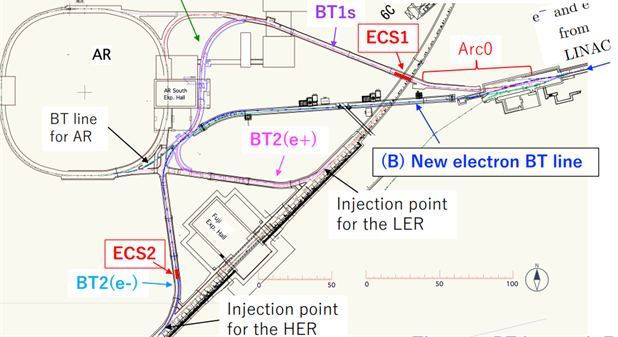}
\caption{A new straighter electron beam transport line in an existing tunnel, as proposed by N.~Iida\protect{\cite{naokoissue25}}.}
\label{fig:newEBT}
\end{figure}

For SuperKEKB, the recommended path towards the design luminosity could include, among other actions:
\begin{enumerate}
\item Clearing the chamber of black stain (vacuum sealant debris), which is presently underway; 
\item developing a strategy for fully correcting and controlling the linear coupling and the chromatic coupling at the interaction point;  
\item eliminating the large sources of local betatron coupling and emittance growth in both electron and positron transfer lines between linac and collider rings, e.g., by installing --- in an existing tunnel --- the proposed new straight e$^-$ beam transport line 
\cite{naokoissue25} to effectively suppress CSR and ISR effects (Fig.~\ref{fig:newEBT}); 
\item further investigating the effect of, plus possibly reducing,  the impact of the transverse damper noise on the SuperKEKB specific luminosity; 
\item reconnecting the BPMs in the final focus with the adjacent magnets and reinforcing the magnet supports; 
\item studying and mitigating the effect of space-charge forces in the LER; 
\item including the effects of nonlinear lattice, space charge, and impedance in the beam-beam simulations in order to explore their combined impact, which has (re-)started;   
\item implementing improved turn-by-turn BPM diagnostics, which will assist both in identifying residual 
optics aberrations at the IP and in understanding the possible degradation of dynamic aperture 
and momentum acceptance with reduced $\beta_y^{\ast}$; and 
\item further optimising the collimation system and reducing losses after injection, e.g., via the new energy compressor, through a longitudinal or hybrid longitudinal-transverse top-up injection scheme, or by suitable pre-collimation of large-amplitude particles 
in the transport lines prior to injection into the collider rings. 
\end{enumerate}
Following this path, SuperKEKB should be able to reach a 
luminosity of $10^{35}$~cm$^{-2}$s$^{-1}$ and beyond.

\section*{Acknowledgements}
I would like to thank Ilya Agapov, 
Andrea Aguirre Polo, 
Vincent Baglin, Philip Bambade, Yunhai Cai, 
Yann Dutheil, Giacomo Broggi, 
Naoko Iida, Jacqueline Keintzel, 
Meng Li, Zhiyuan Li, 
Takashi Mori, Tatsuya Nakada, 
Kazuhito Ohmi, Yukiyoshi Ohnishi, 
Katsunobu Oide, Jack Salvesen, Shinji Terui, 
Rogelio Tomas, and Demin Zhou for valuable discussions, 
helpful information, and for providing some of the 
presented material.    
I am grateful to Patrick Janot and Tor Raubenheimer for encouraging this article, and to Katsunobu Oide, Rogelio Tomas and Patrick Janot for a careful reading of the manuscript.

The work reported here was co-funded 
by the European Union's Horizon-Europe programme under grant no.~101086276 (EAJADE).

\section*{Disclaimer}

{This note is not meant as a substitution for a complete SuperKEKB review, which remains the task of the relevant expert committees. Instead, it is solely aimed at providing the European Strategy Group sufficient input to be able to take an informed decision in December 2025, in view of persistent questions and various rumours according to which the SuperKEKB issues would propagate to FCC-ee and reduce its luminosity ``by a factor 40 or more''. Besides a few pictures specifically produced for this report (e.g.,~Figs.~\ref{fig:ircerr} and \ref{fig:irap}), most, if not all, of the other information reported has already long been publicly disseminated in various forms, and has here been 
synthesised for the aforementioned purpose. 

This report presents the current understanding of the author.  
It does not necessarily reflect the views of the FCC Study Coordination Group, the KEK management, the 
KEKB Accelerator Review Committee, or the EAJADE project. 
Alternative or complementary descriptions and interpretations of the SuperKEKB luminosity performance have been presented elsewhere, e.g., \cite{PhysRevAccelBeams.26.071001,Zhou:2023dhi,gao25}.  
}


\appendix

\section{Other Issues}
Most of the following additional and related issues were raised by K.~Oide, and expanded by the author. 

\subsection{Insufficient beam diagnostics}
\label{sec:diag} 
SuperKEKB lacks turn-by-turn, bunch-by-bunch BPMs, except  for a few out of 450 BPMs per ring. 
Some of the SuperKEKB BPMs (about 50 per ring) 
have a turn-by-turn detection capability 
(albeit with a rather limited resolution, as shown in Table \ref{tab:BPM}),  
but even these cannot record bunch-by-bunch oscillations. 
Therefore, it is not possible to diagnose the beam optics with colliding beams by exciting, or kicking,  
the non-colliding pilot bunches or one of the colliding bunches. As a result, the optics during 
collision at high beam current 
cannot be measured and, hence, also not be corrected.
This lack of even linear optics 
control can be one additional reason for 
luminosity loss at high beam current. 

\begin{table}[htbp]
\caption{Rms precision (noise) of turn-by-turn (TbT) BPM measurements at various storage rings along with other relevant parameters \cite{rogelio25}.}
\label{tab:BPM} 
\centering 
\begin{tabular}{lccccc}
\hline\hline
storage ring &  rms TbT prec. & bunch  &  no.~ & rms bunch  & beam pipe  \\
storage ring &  [$\mu$m] & int.~ [$10^{10}$] & bunches & length [mm] & diameter ($x$,$y$) [mm] \\
\hline
ESRF &  10 & 0.04 & 330 &  &  79, 33 \\
ALBA & 14 & 0.1 & 45 & 6 & 72, 28\\
PETRA III & 22 &  23 & 1--5 & 13.2 & 80, 40 and 94,...\\
LHC & 100 &  1 & 1 & 70 & 49 \\
SuperKEKB LER & 200  & 6 & 1 & & 94 \\
SuperKEKB HER & 125 & 6 & 1 & & 104, 50 \\
\hline\hline
\end{tabular}
\end{table} 

Instead, FCC-ee will be equipped with state-of-the-art 
bunch-by-bunch and turn-by-turn BPMs, as are being used for the LHC and at almost all modern light sources, and as had been pioneered 
at LEP\footnote{The LEP Beam Orbit Measurement (BOM) system, operational from around 1990,  consisted of about 500 BPMs,
each of which could record the horizontal and vertical positions of every bunch for more than 1024 consecutive turns \cite{Borer:1992ys}.},

\subsection{Tunnel floor deformation} 
\label{sec:tunnelfloor} 
\begin{figure}[!ht]%
\centering
\includegraphics[width=0.75\textwidth]{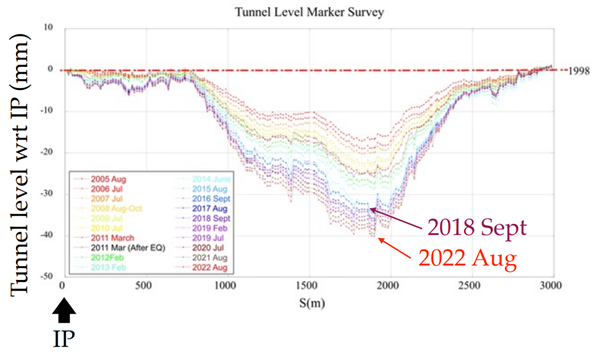}
\caption{Measured year-by-year variation of the (Super)KEKB tunnel 
level with respect to the interaction point,  
presented by M.~Masuzawa in 2022 \cite{arimoto22}.}
\label{fig:motion1}
\end{figure}

\begin{figure}[!ht]%
\centering
\includegraphics[width=0.75\textwidth]{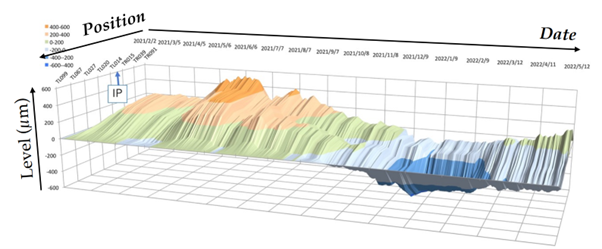}
\caption{Measured seasonal variation of the (Super)KEKB tunnel 
level with respect to the interaction point,  
presented by M.~Masuzawa in 2022 \cite{arimoto22}.}
\label{fig:motion2}
\end{figure}

The SuperKEKB tunnel is not supported by stable rock. Instead, it is floating in the soil. 
A large part of the tunnel is sinking by about 1 mm per year with respect to the interaction point; as is illustrated in Fig.~\ref{fig:motion1}. 
This effect is not taken into account in the usual orbit/optics analysis.  
The vertical tunnel deformation introduces vertical dispersion 
and an uncorrectable vertical emittance. 
Superimposed on the monotonic drift of the vertical tunnel floor level
is a seasonal variation by $\pm$0.5~mm \cite{arimoto22}, 
as is shown in Fig.~\ref{fig:motion2}.  
Due to the long-term and seasonal floor motion, 
also the relative misalignment of the detector and the rest of the accelerator, with consequences described in Section \ref{sec:IR}, could be changing with time.

The FCC-ee collider rings will be located deep underground in the Lake Geneva basin. Any FCC-ee tunnel movements after initial post-construction settling should be much smaller than for SuperKEKB. A stability at least equal to the one of the LEP/LHC tunnel could be assumed for FCC-ee.

\subsection{Limited IR aperture}
\label{sec:irap} 
\begin{figure}[!ht]%
\centering
\includegraphics[width=0.48\textwidth]{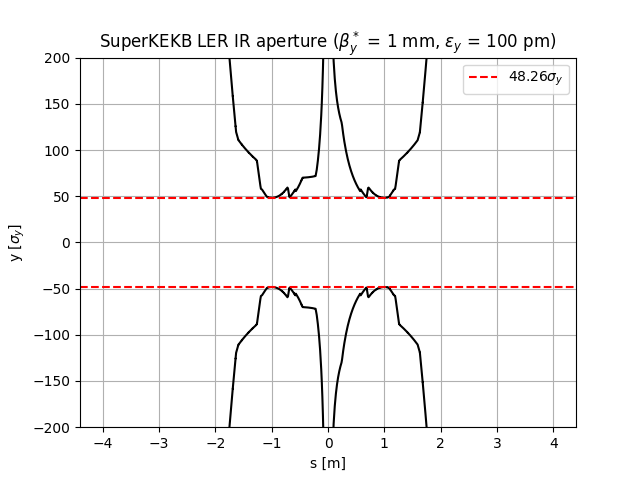}
\includegraphics[width=0.48\textwidth]{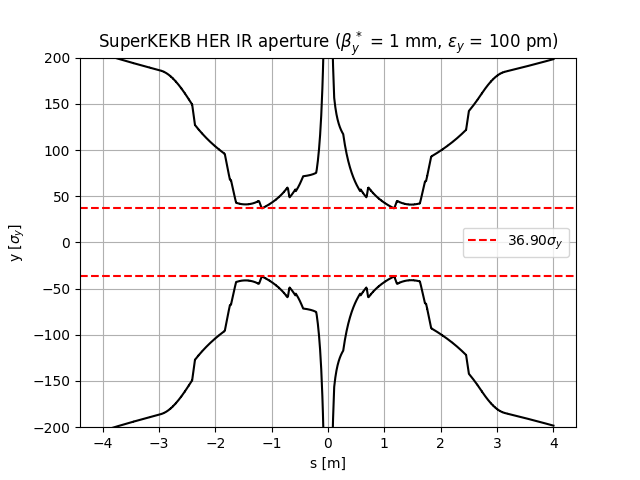}
\caption{Interaction-region physical aperture 
at the SuperKEKB LER (left) and HER (right),
in units of rms vertical beam size, $\sigma_y$, computed for a stored-beam geometric emittance of 100 pm and $\beta_y^{\ast}=1$~mm \cite{broggi} (Courtesy G.~Broggi, 2025).}
\label{fig:irap}
\end{figure}

In conjunction with the much larger than expected
vertical beam emittances,
the mechanical aperture in the SuperKEKB interaction region introduces another limitation. 
Figure \ref{fig:irap} shows the ideal aperture (with zero closed orbit, no misalignments, and zero optics errors) in units of the rms beam size at
$\beta_y^{\ast}=1$~mm, considering a stored-beam
vertical emittance of 100 pm as in Fig.~\ref{fig:emitcol}.   
From Fig.~\ref{fig:irap}, 
the smallest mechanical apertures, under
these optimistic assumptions,
are about 48~$\sigma_y$ in the LER and $\sim$37~$\sigma_y$ for the HER \cite{broggi}. 
To protect the IR aperture
bottlenecks against beam losses,  
the closest vertical collimators of SuperKEKB 
are typically set at about 30~$\sigma_{y}$ in both rings, as confirmed by actual 
collimator settings from the 2024a run \cite{broggi}.
With a normalised vertical injected beam 
emittance of 100~$\mu$m for the LER and
150~$\mu$m for the HER, these collimator settings correspond to
only 2.9~$\sigma$ for the HER and 2.7~$\sigma$ for the LER,
in terms of injected beam sizes. 
At a smaller $\beta_{y}^{\ast}$ of 0.8~mm, the injected beam would be scraped at 2.6~$\sigma_y$ and 2.4~$\sigma_y$, respectively.
At $\beta_{y}^{\ast}=0.3$~mm, the collimators protecting the SuperKEKB IR aperture would sit at only 1.5~$\sigma_y$. 
It is clear that the injected beam emittance should be
reduced before further squeezing $\beta_{y}^{\ast}$.

For the FCC-ee baseline optics, the equilibrium rms vertical emittance in collision is about 2 pm at the Z. 
For this emittance, the aperture bottleneck in the vertical plane is found at the QC1L2 quadrupoles (in the final doublet), with a normalised aperture of 
$\sim 91$~$\sigma_y$ \cite{broggi}.
Also accounting for a possible closed-orbit distortion of up to 250~$\mu$m and a maximum beta-beating of 20\%, the aperture bottleneck in the vertical plane is reduced to about 82~$\sigma_y$.
To protect this aperture,
the closest vertical collimators are set to 50~$\sigma_y$ \cite{broggi}.
For the injected beam, conservatively an emittance of up to 10~pm is considered, which includes a factor 10 margin with respect to the expected booster emittance, in order to allow for any  possible blow up.   
Even under these assumptions,  
the FCC-ee collimator settings, safely protecting the IR apertures,
correspond to 22~$\sigma_y$ 
of the injected beam distribution.  


\subsection{Incomplete beam dynamics}
\label{sec:bd} 
\begin{figure}[!ht]%
\centering
\includegraphics[width=0.65\textwidth]{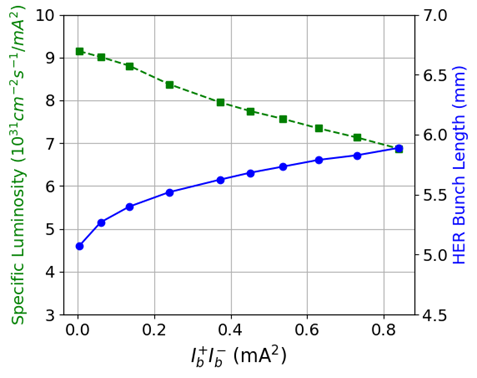}
\caption{Simulated specific luminosity  as a function of bunch-current product (green curve), in APES simulations, including the bunch lengthening predicted from the longitudinal impedance model (blue curve), 
from C.~Lin in 2025, presented by X.~Buffat   \cite{buffat25}.}
\label{fig:zl}
\end{figure}

Some beam dynamics ingredients, such as the effects of the nonlinear lattice, space charge, and impedance, were not routinely 
considered for the SuperKEKB performance estimates. 
A noticeable impact of the nonlinear lattice and space charge on the specific luminosity was pointed out by D.~Zhou 
ten years ago \cite{Zhou:2015cva} 
(see Section \ref{sec:specl}), but left unresolved. 
Bunch lengthening due to the longitudinal impedance will further 
reduce the specific luminosity \cite{PhysRevAccelBeams.26.071001}. 
Figure \ref{fig:zl} indicates a 25\,\% drop in simulated specific luminosity
for a bunch-current-product of 0.8~mA$^2$ (about half the design value) due to the bunch lengthening alone.

Given the many different effects which are each known to noticeably reduce the SuperKEKB specific luminosity with beam current (e.g., the nonlinear lattice, linear coupling, chromatic coupling, higher order aberrations, feedback noise, space charge, longitudinal impedance, ...), it is remarkable  how high a specific luminosity could already be achieved in operation. 

As mentioned in the main text,  at FCC-ee the space charge will be less important due to the higher beam energy and larger bunch length. 
The full nonlinear lattice was used in the luminosity simulations for FCC-ee. Tolerances on errors are being defined based on the targeted performance. The impact of the machine impedances and associated mitigation measures are under study.

\bibliography{skekb}%

\end{document}